\documentclass[12pt]{iopart}
\usepackage{amsfonts}
\begin{document} 

\title[VCS representations, induced representations, and geometric
quantization II]{Vector coherent state representations, induced
  representations, and geometric quantization: \\
  II.\ Vector coherent state representations}

\author{S D Bartlett\dag\ddag, D J Rowe\dag\ and J Repka\S}
\address{\dag\ Department of Physics, University of Toronto, 
  Toronto, Ontario M5S 1A7, Canada}
\address{\ddag\ Department of Physics, Macquarie University, 
  Sydney, New South Wales 2109, Australia}
\address{\S\ Department of Mathematics, University of Toronto, 
  Toronto, Ontario M5S 3G3, Canada}

\ead{stephen.bartlett@mq.edu.au}

\begin{abstract}
  It is shown here and in the preceeding paper~\cite{scalar} that
  vector coherent state theory, the theory of induced representations,
  and geometric quantization provide alternative but equivalent
  quantizations of an algebraic model.  The relationships are useful
  because some constructions are simpler and more natural from one
  perspective than another.  More importantly, each approach suggests
  ways of generalizing its counterparts.  In this paper, we focus on
  the construction of quantum models for algebraic systems with
  intrinsic degrees of freedom.  Semi-classical partial
  quantizations, for which only the intrinsic degrees of freedom are
  quantized, arise naturally out of this construction.  The
  quantization of the $SU(3)$ and rigid rotor models are considered as
  examples.
\end{abstract}

\submitto{\JPA}

\section{Introduction} 

Quantizing a classical model is a difficult problem in general.  The
theory of geometric quantization (GQ)~\cite{gq} provides a general and
powerful framework for the quantization of a wide variety of classical
systems, but due to its formidable mathematical language it is
inaccessible to most physicists.  We show here and in the preceding
paper~\cite{scalar} that the useful and physically-motivated theory
of coherent state representations~\cite{per86,Klauder} provides a
natural language for describing the techniques of GQ.
In~\cite{scalar}, it was shown that scalar coherent state theory
yields three categories of representations for the spectrum generating
algebra (SGA) of an algebraic model: classical realizations,
prequantization, and the irreducible representations of quantization.
This paper generalizes the results of \cite{scalar} to vector-valued
coherent state representations.

While it is often possible to induce representations of a Lie algebra
from a one-dimensional irrep of some subalgebra (as in the standard
coherent state construction), it is generally more economical and
effective to induce from a known multi-dimensional irrep of a larger
subalgebra.  The amount of work is then minimized by capitalizing on
information that is already available, and leads to a useful physical
interpretation of some degrees of freedom of a model system as
intrinsic.  For example, using the method of induced representations,
Wigner~\cite{wigner} found irreps of the Poincar\'e group
corresponding to quantizations of particles with intrinsic spin.  Such
intrinsic degrees of freedom are often regarded as having quantal
origins.  It will be seen that they have classical counterparts and
that the general theory of induced representations, when developed
within the framework of vector coherent state (VCS)
theory~\cite{vcs,rowe91}, has a natural expression in the language of
geometric quantization.

\section{Classical representations with intrinsic degrees of freedom} 
\label{sect:4.classical}

Let $T$ be an abstract (possibly projective) unitary representation of
a dynamical group $G$ on a Hilbert space $\mathbb{H}$.  As in the
scalar theory, $T$ need not be specified precisely; it could be, for
example, a regular representation, or a Weil representation on a
many-particle Hilbert space.  Corresponding to any normalized state
$|0\rangle\in \mathbb{H}$ there is a coadjoint orbit
\begin{equation}
  \label{eq:CoadjointOrbit}
  \mathcal{O}_\rho= \{ \rho_g ; g\in G\}
\end{equation}
of densities defined  by
\begin{equation}
  \label{eq:Densities}
  \rho_g(A) = \langle0|\hat A(g)|0\rangle \,,
\end{equation}
where $\hat A= T(A)$ and $\hat A(g) = T(g) \hat AT(g^{-1})$.  Let
$H_\rho \subset G$ be the isotropy subgroup of $\mathcal{O}_\rho$ at
$\rho$; the orbit $\mathcal{O}_\rho \simeq H_\rho \backslash G$ is
known to be symplectic and can be regarded as a classical phase space.
Moreover, a classical representation ${\cal A}$ of an element $A \in
\mathfrak{g}$, the Lie algebra of $G$, is given as a function on
$\mathcal{O}_\rho$ by $\mathcal{A}(g)= \rho_g(A)$ (for
details, see~\cite{scalar}).

Let $H\supseteq H_\rho$ be some other subgroup. It may be convenient
to choose $H$ such that $H\backslash G$ is also symplectic, but this
condition is not necessary.  The phase space $\mathcal{O}_\rho \simeq
H_\rho\backslash G$ may then be viewed as a $H_\rho\backslash G\to
H\backslash G$ fibre bundle with typical fibre $H_\rho\backslash H$.
When $H$ is set equal to $H_\rho$, as in scalar coherent state theory,
the fibres become trivial.  A specification of $H$ that contains
$H_\rho$ as a proper subgroup, in vector coherent state theory,
corresponds to regarding some degrees of freedom of $G$ as intrinsic,
i.e., as gauge degrees of freedom.  We refer to $H$ as the
\emph{intrinsic symmetry group}.

Viewing the classical phase space as a smaller space with intrinsic
degrees of freedom in this way does not change a classical
representation in principle.  However, it gives a new perspective and
leads to new quantization procedures.  Starting with a density
$\rho\in \mathfrak{g}^*$, the classical phase space $\mathcal{O}_\rho$
is generated in two steps.  The first step generates the coadjoint
orbit $H_\rho \backslash H$ of the subgroup $H\subset G$ as the set of
densities $\{\rho_\alpha ; \alpha \in H\}$.  This set is then regarded
as the fibre of a bundle over the point $H$ of the space $H\backslash
G$.  The second step defines the fibre over an arbitrary point $Hg$ of
$H\backslash G$ as the set $\{\rho_{\alpha g} ; \alpha \in H\}$. The
classical function $\mathcal{A}$ on $G$ representing an element $A\in
\mathfrak{g}$, defined as having values $\mathcal{A}(g) =\rho(A(g))$,
with $A(g) = {\rm Ad}_g(A)$ ($=gAg^{-1}$ for a matrix group) is then
seen as being $H$-equivariant, i.e., it satisfies the equation
\begin{equation}
  \mathcal{A}(\alpha g) =  \rho_\alpha(A(g)), \quad
  \forall\, \alpha\in H \, .
\end{equation}
When $H=H_\rho$, the fibres are trivial and this equivariance
condition reduces to the invariance condition $\mathcal{A}(\alpha g)
=\mathcal{A}(g)$ for $\alpha\in H$.

As an example, consider a particle moving in a three-dimensional
Euclidean space.  If the particle has intrinsic spin, it is
appropriate to take as SGA the semidirect sum of $hw(3)$, a
Heisenberg-Weyl algebra, and $su(2)$ with basis
$\{\hat{q}_i,\hat{p}_i,\hat{I},\hat{J}_i; i = 1,2,3\}$ and commutation
relations
\begin{equation}
  \label{eq:hw(3)su(2)LieAlgebra}
  \eqalign{
  [\hat{q}_i,\hat{p}_j] = \rmi\hbar \delta_{ij}\hat{I} \, , &
  [\hat{J}_i,\hat{q}_j] = \rmi \hbar \sum_k \varepsilon_{ijk} \hat{q}_k \,
  , \\
  [\hat{J}_i,\hat{J}_j] = \rmi \sum_k \varepsilon_{ijk} \hat{J}_k \, , \qquad&
  [\hat{J}_i,\hat{p}_j] = \rmi \hbar \sum_k \varepsilon_{ijk} \hat{p}_k
  \, .
  }
\end{equation}
We suppose these Lie algebra elements act via a representation $T$ as
Hermitian operators on some Hilbert space $\mathbb{H}$.  Let
$|0\rangle \in \mathbb{H}$ be a state with expectation values
\begin{equation}
  \eqalign{
  \langle 0|\hat I|0\rangle = 1 \, ,\qquad \langle 0|\hat J_3|0\rangle
  = M \, , \\
  \langle 0|\hat q_i|0\rangle = \langle 0|\hat p_i|0\rangle = 
  \langle 0|\hat J_1|0\rangle = \langle 0|\hat J_2|0\rangle = 0 \, .
  }
\end{equation}
An $[HW(3)]SU(2)$ group element can be parameterized
\begin{equation}
  T(g(v,q,p)) =  T(v) \,
  \rme^{-\frac{\rmi}{\hbar}\sum_i p_i \hat{q}_i } 
  \rme^{\frac{\rmi}{\hbar}\sum_i q_i
  \hat{p}_i} \, ,
\end{equation}
with $v$ a $U(2)$ group element.  For $M \neq 0$, the isotropy
subgroup $H_\rho$ of the density defined by $\rho(\hat A) = \langle
0|\hat A|0\rangle$ is the group $H_\rho \simeq U(1)\times U(1)$ with
infinitesimal generators $\{ \hat I, \hat J_3\}$.  Thus, with
$\rho(\hat A) = \langle 0|\hat A|0\rangle$ and $g=g(v, q,p)$, the
classical representation of the observables
$\{\hat{q}_i,\hat{p}_i,\hat{I},\hat{J}_i\}$ is given by the functions
$\{ \mathcal{Q}_i, \mathcal{P}_i, \mathcal{I}, \mathcal{J}_i\}$ with
\begin{equation}
  \eqalign{
  \mathcal{Q}_i(g) &=  \rho(\hat q_i(g))  = q_i \, , \\
  \mathcal{P}_i(g) &=   \rho(\hat p_i(g)) = p_i \, , \\
  \mathcal{J}_i(g) &=   \rho(\hat J_i(g)) =
  \mathcal{S}_i(v) + (q_jp_k-q_kp_j) \, ,
  }
\end{equation}
where $\mathcal{S}_i$, a function over $SU(2)$, represents the
intrinsic spin of the particle.  The functions of this classical
representation can be regarded as functions over $H_\rho \backslash
[HW(3)]SU(2)$.  However, they are more usefully represented as
functions over the classical $(p-q)$ phase space, $U(2)\backslash
[HW(3)]SU(2)\simeq U(1)\backslash HW(3)$, with intrinsic spin degrees
of freedom defined by a choice of intrinsic symmetry group $H= U(2)$.

In the following sections, we show that VCS theory produces three
categories of quantization of an algebraic model with intrinsic
degrees of freedom: (i) semi-classical partial quantizations for
which only the intrinsic degrees of freedom are quantized; (ii)
unitary reducible representations that have the form of a
prequantization; and (iii) unitary irreps of a full quantization,
equivalent to those obtained by GQ but with an additional fibre
structure encompassing the intrinsic degrees of freedom.  Each
category of representation is a natural extension of the scalar
theory.

\section{Semi-classical partial quantizations} \label{sec:Partial}

A \emph{partial quantization} is a representation in which only the
intrinsic degrees of freedom are quantized and the extrinsic degrees
of freedom are represented classically.

To be specific, suppose that $M$ is an irreducible unitary
representation of an intrinsic symmetry group $H$ on a
finite-dimensional (intrinsic) Hilbert space $U$.  Then a partial
quantization is obtained by replacing the classical phase space, seen
as a $H_\rho\backslash G\to H\backslash G$ bundle with typical fibre
$H_\rho\backslash H$, by a semi-classical state space $B$ with the
geometric structure of a fibre bundle associated to the principal $G
\to H\backslash G$ bundle by the representation $M$ of $H$.  A
semi-classical state of the system, corresponding to a point in $B$,
is then a state vector in the intrinsic Hilbert space over a point in
the classical $H\backslash G$ phase space.

Semi-classical representations result when the scalar coherent state
construction of a classical representation is generalized to a VCS
construction.  As a prelude to defining $H$ and $M$, we start with a
finite-dimensional subspace $U\subset\mathbb{H}$ of the Hilbert space
for an abstract unitary representation $T$ of the dynamical group $G$.
Denote by $E$ the natural embedding $E : U\to\mathbb{H}$.  There is
then a system $\{ U(g) ; g\in G\}$ of coherent state subspaces in
$\mathbb{H}$ defined by
\begin{equation}
  U(g) = \{ |\psi(g)\rangle = T(g^{-1})|\psi\rangle ; |\psi\rangle\in
  E(U)\} \,.
\end{equation}
Let $\Pi$ denote the projection of $\mathbb{H}$ to $U$ relative to the
inner product on $\mathbb{H}$.  Then the subspace $U\subset\mathbb{H}$
defines a map $\hat \rho :\mathfrak{g}\to GL(U)$ from the Lie algebra
$\mathfrak{g}$ to the linear transformations of $U$ by
\begin{equation}
  \hat \rho(A) = \Pi \hat A E \,, \quad \forall\, A\in \mathfrak{g} \,.
\end{equation}
In the special case that $U$ is one-dimensional and spanned by a
state of unit norm $|0\rangle$, $\hat \rho$ reduces to a scalar
density and acts on an arbitrary vector $|\psi\rangle\in U$ by scalar
multiplication, i.e., $\hat\rho(A)|\psi\rangle = |\psi\rangle \langle
0| \hat A |0\rangle$.  Thus, the above definition of $\hat\rho$
generalizes the concept of a density $\rho: \mathfrak{g} \to
\mathbb{R}$ to a map $\hat \rho :\mathfrak{g}\to GL(U)$; we therefore
refer to $\hat \rho$ as a \emph{semi-classical density}.  The set of
such semi-classical densities
\begin{equation}
  \mathcal{O}_{\hat \rho} = \{ \hat\rho_g ; g\in G\} \, ,
\end{equation}
defined by
\begin{equation}
  \hat\rho_g(A) = \hat\rho(A(g)) \, ,
\end{equation}
is then a natural generalization of a coadjoint orbit.

The orbit $\mathcal{O}_{\hat\rho}$ has the structure of a fibre bundle
over $H\backslash G$, where $H$ is a subgroup of $G$ with Lie
algebra
\begin{equation} \label{eq:3.12}
  \mathfrak{h} = \{ A\in \mathfrak{g} |\, \hat\rho ([A,X]) = 
  [\hat\rho(A),\hat\rho(X)] \,,\;\forall\, X\in \mathfrak{g}\} \,.
\end{equation}
With this definition, $\mathfrak{h}$ is a subalgebra of $\mathfrak{g}$
for which the restriction of $\hat\rho$ to $\mathfrak{h}\subset
\mathfrak{g}$ is a representation.  Let $M$ be an extension of this
representation to the group $H$ such that
\begin{equation}
  \rmi\frac{\rmd}{\rmd t} M(\rme^{-\rmi At})\Big|_{t=0} = M(A) \equiv
  \hat\rho (A) \,,\quad\forall\, A\in \mathfrak{h}\,,
\end{equation}
and
\begin{equation}
  \hat\rho(X(hg)) =
  M(h) \hat\rho(X(g))M(h^{-1}) \,, \quad \forall\ h\in H\,,\ X\in
  \mathfrak{g}\,.
\end{equation}
The elements of $\mathcal{O}_{\hat\rho}$ then satisfy the
$H$-equivariance condition
\begin{equation}
  \label{eq:HAction}
  \hat\rho_{hg}  =M(h) \hat\rho_gM(h^{-1})
  \,, \quad \forall\  h\in H\,,
\end{equation}
and $\mathcal{O}_{\hat\rho}$ is interpreted as a fibre bundle over
$H\backslash G$ associated to the principal $G\to H\backslash G$
bundle by the action~(\ref{eq:HAction}).  The $H$-equivariance
condition is a generalization of the $H$-invariance condition for the
scalar densities of a standard coadjoint orbit $\mathcal{O}_{\rho}\sim
H_\rho\backslash G$;
\begin{equation}
  \rho_{hg} = \chi(h) \rho_g \chi(h^{-1}) = \rho_g \,, \quad \forall\
  h\in H_\rho\,.
\end{equation}

It is interesting to note that the representation $M$ of
$\mathfrak{h}$ and $H$, defined by
\begin{equation}
  M : A\to \hat\rho (A) \,,\quad \forall\ A\in \mathfrak{h} \,,
\end{equation}
is generally not a subrepresentation of the restriction of the
representation $T$ to $\mathfrak{h}\subset\mathfrak{g}$.  The parallel
of this observation was obvious for the abelian scalar representation
\begin{equation}
  \chi : A\to \langle 0| T(A) |0\rangle \,, \quad A\in
  \mathfrak{h}_\rho\,,
\end{equation}
but it is less obvious that multidimensional representations that are
not subrepresentations exist.  However, they are known for some Lie
algebras and are described as \emph{embedded}
representations~\cite{rowerochrepk88}.

Note also that the representation $M$ could be reducible. However,
although it is not essential, we shall assume in the following that
the subspace $U\subset\mathbb{H}$ is chosen in such a way that it is
irreducible.

The semi-classical density now defines a semi-classical
representation of $\mathfrak{g}$ in which an element $A\in
\mathfrak{g}$ is mapped to an operator-valued function
$\hat{\mathcal{A}}$ over $G$ having values
\begin{equation}
  \label{eq:hatAfunction}
  \hat{\mathcal{A}}(g) = \hat\rho_g(A)= \hat\rho(A(g)) \, ,  
\end{equation}
in $GL(U)$, which satisfies the equivariance relationship
\begin{equation}
  \hat{\mathcal{A}}(hg) = M(h)
  \hat{\mathcal{A}}(g)M(h^{-1})\,,\quad \forall\, h\in H \, .
\end{equation}
The Poisson bracket for this representation is defined by
\begin{equation}
  \label{eq:semiclassLieBracket}
  \{\hat{\mathcal{A}},\hat{\mathcal{B}}\}(g) = - \frac{\rmi}{\hbar}
  \hat\rho ([A(g),B(g)]) \, .
\end{equation}

Let $\{ A_i \}$ be a basis for $\mathfrak{h}$ and $\{A_\nu \}$ a
complementary set that completes a basis for $\mathfrak{g}$.   From the
expansion 
\begin{equation}
  A(g) = \sum_i A^i(g) A_i + \sum_\nu A^\nu(g) A_\nu \, ,
  \label{eq:expansion}
\end{equation}
it follows that
\begin{equation}
  \hat{\mathcal{A}}(g) = \sum_i A^i(g) M(A_i) + \sum_\nu A^\nu(g)
  \hat\rho(A_\nu)\, ,
  \label{eq:expandAhat}
\end{equation}
and that
\begin{equation}
  \label{eq:SemiclassPoissonBracketDecomposition}
  \{\hat{\mathcal{A}},\hat{\mathcal{B}}\}(g) =  - \frac{\rmi}{\hbar}
  [\hat{\mathcal{A}}(g),\hat{\mathcal{B}}(g)] 
  + \sum_{\mu\nu} A^\mu (g) \hat\Omega_{\mu\nu} B^\nu(g) \, ,
\end{equation}
where
\begin{equation}
  \hat\Omega_{\mu\nu} = - \frac{\rmi}{\hbar}\Bigl( \hat\rho\bigl([A_\mu,
  A_\nu]\bigr) - [\hat\rho(A_\mu),\hat\rho(A_\nu)]\Bigr) \, .
\end{equation}

Following standard terminology, it is convenient to characterize the
decomposition of a Lie algebra element into a vertical component (an
element of $\mathfrak{h}$) and a complementary (horizontal) component,
as a choice of gauge.  Thus, a gauge is defined by a projection
$\mathfrak{g}\to \mathfrak{h}; A(g) \mapsto\sum_i A^i(g) A_i$.  It is
then notable that the second term of
equation~(\ref{eq:SemiclassPoissonBracketDecomposition}) is gauge
independent.  This independence follows from the definition of
$\mathfrak{h}$, equation~(\ref{eq:3.12}), which implies that
\begin{equation}
  \sum_{\mu\nu} A^\mu (g) \hat\Omega_{\mu\nu} B^\nu(g)= -
  \frac{\rmi}{\hbar} \Bigl( \hat\rho\bigl([A(g),B(g)]\bigr) -
  [\hat\rho(A(g)),\hat\rho(B(g))]\Bigr) \, .
\end{equation}
Consequently, as shown in the appendix, the semi-classical Poisson
bracket of equation~(\ref{eq:SemiclassPoissonBracketDecomposition})
has a manifestly covariant expression
\begin{equation}
  \label{eq:SemiclassPoissonBracketDecomposition2}
  \rmi\hbar \{\hat{\mathcal{A}},\hat{\mathcal{B}}\}(g) = 
  [\hat{\mathcal{A}}(g),\hat{\mathcal{B}}(g)] + 
  \rmi\hbar \hat\Omega
  \bigl(X_{\hat{\mathcal{A}}}(g),X_{\hat{\mathcal{B}}}(g) \bigr) \,,
\end{equation}
where $X_{\hat{\mathcal{A}}}$ is a Hamiltonian vector field generated by
$\hat{\mathcal{A}}$ and $\hat\Omega$ is a curvature tensor for the
semi-classical phase space  (both of which are defined in
the appendix).

While for formal purposes it is convenient to express a classical
representation by functions over the group $G$, it is generally more
useful, in practical applications, to represent them as functions over
a suitable set of $H\backslash G$ coset representatives. Recall that a
set of coset representatives $K=\{ k(g)\in Hg ; g\in G\}$ defines a
factorization $g = h(g) k(g)$, with $h(g)\in H$, of every $g\in G$.
Hence, it follows from the identity
\begin{equation}
  \hat{\mathcal{A}}(h(g)k(g)) =  M(h(g)) 
  \hat{\mathcal{A}}(k(g))M(h^{-1}(g)) \,, 
\end{equation}
that, given the representation $M$, the restriction of
$\hat{\mathcal{A}}$ to the subset $K\subset G$ is sufficient to
uniquely define $\hat{\mathcal{A}}$.  Moreover, the Poisson bracket of
two such functions is given directly in terms of this restriction by
\begin{equation}
  \{\hat{\mathcal{A}},\hat{\mathcal{B}}\}(k) = - \frac{\rmi}{\hbar}
  \hat\rho \bigl([A(k),B(k)]\bigr) \, , \quad \forall\, k\in K\,.
\end{equation}
Often it is convenient to consider factorizations of the type $g =
h(g)k(g)$ with $h(g) \in H^c$ and $k(g) \in K$, where $K$ is a subset
of $H^c\backslash G^c$ coset representatives and $H^c$ and $G^c$ are
the complex extensions of $H$ and $G$, respectively.  The
semi-classical representation is then by operator-valued functions
on $K$.

As an illustration of partial quantization, suppose the intrinsic spin
observables of a particle in a three-dimensional Euclidean space,
cf.\ section~\ref{sect:4.classical}, are described quantally by a
finite-dimensional irrep $M $ of the $u(2)$ intrinsic symmetry
algebra.  Let $\{ \xi_{sm}; m=-s,\ldots,s \}$ be an orthonormal basis
for the Hilbert space $U$ of this irrep.  Let $E : U\to \mathbb{H} ;\ 
\xi_{sm} \mapsto |sm\rangle$ be an embedding of $U$ as an
$su(2)$-invariant subspace of $\mathbb{H}$ such that
\begin{equation}
  \langle sm |\hat q_i |sn\rangle =\langle sm |\hat p_i |sn\rangle = 0 \,,
  \qquad \langle sm |\hat I |sn\rangle = \delta_{mn} \,,
\end{equation}
and define 
\begin{equation}
  \hat\rho( A) = \sum_{mn} \xi_{sm} \langle sm |\hat A |sn\rangle \,
  \xi^\dagger_{sn} \,, \quad \forall\ A\in \frak{g}\,,
\end{equation}
with the understanding that $\xi^\dagger_{sn} \cdot \xi_{sm}=
\delta_{mn}$.  The semi-classical representation of the
$[hw(3)]su(2)$ algebra can be defined on the coset space $U(2)
\backslash [HW(3)]SU(2)$ (i.e., the $p-q$ plane) as
\begin{equation}
  \eqalign{
  \hat{\mathcal{Q}}_i(p,q) &=  \hat{\rho}(\hat{q}_i(g))  = 
   q_i \hat{\mathcal{I}}\,  ,  \\
  \hat{\mathcal{P}}_i(p,q) &=  \hat{\rho}(\hat{p}_i(g)) =
   p_i\hat{\mathcal{I}} \, , \\
  \hat{\mathcal{J}}_i(p,q) &=  \hat{\rho}(\hat{J}_i(g)) =
  \hat{\mathcal{S}}_i+ \hat{\mathcal{L}}_i(p,q) \, ,
  } 
\end{equation}
where
\begin{equation}
  \hat{\mathcal{S}}_i = \hat\rho(\hat J_i) \,, \qquad 
  \hat{\mathcal{L}}_i(p,q) = (q_jp_k-q_kp_j) \hat{\mathcal{I}} 
\end{equation}
are the spin and orbital angular momenta, respectively, and
$\hat{\mathcal{I}} = \hat\rho(\hat{I})$ is the identity operator on
$U$.  The quantal part of the Lie bracket for these semi-classical
observables is now given by
\begin{equation}
  \eqalign{
  {[}\hat{\mathcal{Q}}_i(p,q),\hat{\mathcal{P}}_i(p,q)] = 
  [\hat{\mathcal{J}}_i(p,q),\hat{\mathcal{Q}}_i(p,q)] =
  [\hat{\mathcal{J}}_i(p,q),\hat{\mathcal{P}}_i(p,q)] =0 \,, \\
  {[}\hat{\mathcal{J}}_i(p,q),\hat{\mathcal{J}}_j(p,q){]} = \rmi\hbar
  \hat{\mathcal{S}}_k \,,
  }
\end{equation}
and the classical part by 
\begin{equation}
  \eqalign{
  \rmi\hbar\hat\Omega \bigl(
  X_{\hat{\mathcal{Q}}_i},X_{\hat{\mathcal{P}}_i}\bigr)(p,q) =
  \rmi\hbar\hat{\mathcal{I}}  \,,
  & \rmi\hbar\hat\Omega 
  \bigl(X_{\hat{\mathcal{J}}_i},X_{\hat{\mathcal{Q}}_j}\bigr)(p,q) = 
  \rmi\hbar\hat{\mathcal{Q}}_k(p,q) \,, \\
  \rmi\hbar\hat\Omega \bigl(X_{\hat{\mathcal{J}}_i},
  X_{\hat{\mathcal{P}}_j}\bigr)(p,q) =  
  \rmi\hbar\hat{\mathcal{P}}_k(p,q) \,, \quad&
  \rmi\hbar\hat\Omega \bigl(X_{\hat{\mathcal{J}}_i},
  X_{\hat{\mathcal{J}}_j}\bigr)(p,q) = 
  \rmi\hbar\hat{\mathcal{L}}_k(p,q) \,. 
  }
\end{equation}
Together, these parts lead to a semi-classical representation of 
$[hw(3)]su(2)$ with Poisson bracket given by
equation~(\ref{eq:SemiclassPoissonBracketDecomposition2}).

Such semi-classical representations not only provide a useful and
insightful first step in the quantization of a complex system, they
are also of considerable physical interest in their own right.  For
example, in many situations involving macroscopic degrees of freedom,
a classical description of the dynamics is more than adequate.
However, macroscopic systems can also have microscopic intrinsic
structures for which quantum mechanics is essential.  For example, it
may be appropriate to quantize the intrinsic dynamics of a heavy
molecule but to describe its center-of-mass motions classically.
The scattering of a heavy ion by a nucleus might be another example.
The existence of corresponding partial quantizations of their spectrum
generating algebras is therefore a potentially powerful tool in their
analysis.

\section{VCS induced representations as prequantization}
\label{sect:VCSinduced}

A VCS representation can be constructed in the form of a
prequantization.  It will be convenient to say that an irrep $M$ of $H
\subset G$ is \emph{contained} in a (possibly projective)
representation $T$ of $G$ if $M$ appears in either a direct sum or
direct integral decomposition of $T_H$, where $T_H$ is the restriction
of $T$ to $H\subset G$.  We then say that a semi-classical
representation of $\mathfrak{g}$, defined by an irrep $M$ of a compact
intrinsic symmetry group $H\subset G$, is quantizable if $M$ is
contained in some unitary representation $T$ of the group $G$ on a
Hilbert space $\mathbb{H}$.  It follows, by Schur's lemma, that if $M$
is quantizable there exists a non-vanishing $H$-intertwining
operator $\Pi :\mathbb{H}_D\to U$, from a dense subspace of
$\mathbb{H}$ to $U$, the carrier space of $M$, such that
\begin{equation}
  \label{eq:IntertwiningOperatorProperty}
  \Pi T(h) = M(h) \Pi, \quad \forall\ h
  \in H \, .
\end{equation}

Given an abstract unitary representation $T$ of $G$ and such an
$H$-intertwining operator, a VCS wave function $\Psi$ is defined over
$G$~\cite{rowe91} for every $|\Psi\rangle\in\mathbb{H}_D$ by
\begin{equation}
  \label{eq:VCSwfuns}
  \Psi(g) = \Pi T(g) |\Psi\rangle \, , 
  \quad \forall\ g\in G \, . 
\end{equation}
It follows from the definition of $\Pi$ that
\begin{equation}
  \Psi(hg) = M(h) \Psi(g) \, ,\quad \forall\ h\in H \, . 
  \label{eq:LeftAction}
\end{equation}
A VCS representation $\Gamma$ of the group $G$ induced from the
representation $M$ of the subgroup $H\subset G$, is then defined by
\begin{equation}
  [\Gamma(g') \Psi](g) = \Psi(gg') \, , \quad g'\in G \, . 
  \label{eq:VCSaction}
\end{equation}
Equations~(\ref{eq:LeftAction}) and (\ref{eq:VCSaction}), of which the
scalar coherent state representations are special cases, are the basic
equations of all inducing constructions.

For example, suppose $M$ is a representation of $H$ on a Hilbert space
$U$ with orthonormal basis $\{ \xi_m\}$ and $E : U\to \mathbb{H} \,;\;
\xi_m\mapsto |m\rangle$ is an embedding of $U$ as an $H$-invariant
subspace $E(U)$ in $\mathbb{H}$.  Then a suitable intertwining
operator is defined by
\begin{equation}
  \Pi = \sum_m \xi_m \langle m | \, ,
\end{equation}
and vector coherent state wave functions are expressed
\begin{equation}
  \Psi(g) = \sum_m \xi_m \langle m |T(g) |\Psi\rangle \,.
\end{equation}

In principle, the Hilbert space of VCS wave functions is determined by
the map (\ref{eq:VCSwfuns}) from $\mathbb{H}_D$ to VCS wave functions;
the inner product can be inferred as in section 3.4 of the preceding
paper~\cite{scalar}.  Many VCS Hilbert spaces are possible depending
on the choice of $T$ and the embedding $E$.  For example, as discussed
briefly in section~\ref{subsec:VCSHilbert}, if $T$ is the regular
representation of the group $G$ and $E$ has no special properties,
then $\Gamma$ is the representation of $G$ induced from the
representation $M$ of a subgroup $H\subset G$ in the standard theory
of induced representations.  This representation is known to be
reducible in general and, as we now show, it is a natural
generalization of a prequantization.  However, the embedding $E$ can
also be chosen such that the VCS representation is a subrepresentation
of the standard induced representation.  It is shown in the following
section that it can even be chosen such that the VCS representation is
irreducible.

Following the construction of the scalar coherent state
representations, the general inducing construction defines a
representation of the Lie algebra $\mathfrak{g}$ by
\begin{equation} 
  \label{eq:4.defn}
  [\Gamma(A) \Psi](g) = \Pi T(g)T(A)|\Psi\rangle
  = \Psi(A(g)g) \, , \quad A\in \mathfrak{g} \,,
\end{equation}
where $\Psi(Ag)$ is defined generally, for any $A\in
\mathfrak{g}$ by
\begin{equation}
  \Psi(Ag) = \rmi\frac{\rmd}{\rmd t}\Psi\bigl(\rme^{-\rmi tA}g\bigr)
  \Big|_{t=0} \,. 
  \label{eq:4.psiAg}
\end{equation} 

For a given choice of gauge, defined by a basis $\{ A_i\}$ for
$\mathfrak{h}$ and a complementary set $\{ A_\nu\}$ to complete a
basis for $\mathfrak{g}$, the expansion of $A(g)$ given by
equation~(\ref{eq:expansion}) leads to the explicit expression
\begin{equation} 
  \label{eq:4.42}
  [\Gamma(A) \Psi](g) = \sum_i A^i(g) M(A_i) \Psi(g) + \rmi\hbar \sum_\nu
  A^\nu(g) [\partial_\nu \Psi](g) \,,
\end{equation}
where
\begin{equation}
  [\partial_\nu \Psi](g) = \frac{\partial}{\partial x^\nu} \Psi\bigl(
  \rme^{-\frac{\rmi}{\hbar}  \sum_\mu x^\mu A_\mu} g\bigr)\Big|_{x=0} \,.
\end{equation}
Note that this generalization of a scalar coherent state
representation is achieved simply by replacing the one-dimensional
representation $\chi$ of the intrinsic symmetry group by the
multidimensional representation $M$.

Like its scalar counterpart, the representation $\Gamma$ can be
expressed in the covariant form of a prequantization.  From
equation~(\ref{eq:expandAhat}), we have
\begin{equation}
  \sum_i A^i(g) M(A_i) = \hat{\mathcal{A}}(g) - \sum_\nu A^\nu (g) 
  \hat{\rho}(A_{\nu}) \,.
\end{equation}
Equation~(\ref{eq:4.42}) then becomes
\begin{equation} 
  \label{eq:4.4}
  [\Gamma(A) \Psi](g) = \hat{\mathcal{A}}(g) \Psi(g) + \rmi\hbar \sum_\nu
  A^\nu(g) [\nabla_\nu \Psi](g) \,,
\end{equation}
where
\begin{equation}  
  \nabla_\nu = \partial_\nu + \frac{\rmi}{\hbar} \hat\rho(A_\nu) \,.
\end{equation}
The first term, $\hat{\mathcal{A}}(g) \Psi(g)$, of equation~(\ref{eq:4.4}) is
manifestly covariant.  Moreover, from the definition (\ref{eq:4.defn}),
the second term is identical to 
\begin{equation}  
  \rmi\hbar[\nabla_A\Psi](g) = \Psi(A(g)g) - \hat\rho (A(g)) \Psi(g) \,.
\end{equation}
where
\begin{equation}  \label{eq:covderiv}
  [\nabla_A \Psi](g) = \sum_\nu A^\nu (g) [\nabla_\nu\Psi](g) \,,
\end{equation}
Thus, it too is covariant.

It is shown in the appendix that $\nabla_A$ is identical to the
covariant derivative $\nabla_{X_{\hat{\mathcal{A}}}}$ in the direction
of the vector field $X_{\hat{\mathcal{A}}}$ and is expressed in a
particular gauge as a sum
\begin{equation}
  \nabla_A = \nabla_{X_{\hat{\mathcal{A}}}} =X_{\hat{\mathcal{A}}}
  +\frac{\rmi}{\hbar}\hat\theta(X_{\hat{\mathcal{A}}}) \, ,
\end{equation}
where $X_{\hat{\mathcal{A}}}$ is a Hamiltonian vector field generated
by $\hat{\mathcal{A}}$ and $\hat\theta$ is a one-form.  It is also
shown that the curvature $\hat\Omega$ of the semi-classical phase
space is the covariant exterior derivative of $\hat\theta$ given by
\begin{equation}
  \hat\Omega(X_{\hat{\mathcal{A}}},X_{\hat{\mathcal{B}}}) = 
  \rmd\hat\theta(X_{\hat{\mathcal{A}}},X_{\hat{\mathcal{B}}}) -
  [\hat\theta(X_{\hat{\mathcal{A}}}),\hat\theta(X_{\hat{\mathcal{B}}})]
  \,.
\end{equation}

Thus, the VCS representation $\Gamma(A)$ of an element
$A\in\mathfrak{g}$ is expressed
\begin{equation}
  \label{eq:VCSPrequantization}
  \Gamma(A) = \hat{\mathcal{A}} + \rmi\hbar \nabla_{X_{\hat{\mathcal{A}}}} \,,
\end{equation}
and is seen as a natural generalization of prequantization to include
intrinsic degrees of freedom.

As for semi-classical observables, it is generally more useful
to express VCS wave functions as functions over a suitable set of
$H\backslash G$ coset representatives.  Thus, with a set of coset
representatives $K=\{ k(g)\in Hg ;\, g\in G\}$, it follows from the
identity
\begin{equation}
  \Psi(g) = \Psi(h(g) k(g)) = M(h(g)) \Psi(k(g)) \,,
\end{equation}
that, given $M$, the restriction of $\Psi$ to the subset $K\subset G$
is sufficient to uniquely define $\Psi$.  VCS wave functions can also
be defined over a subset of $H^c\backslash G^c$ coset representatives
$K^c$ by a factorization $g = h(g) k(g)$, with $h(g)\in H^c, k(g)\in
K^c$, of every $g\in G$.

For the example of a particle with intrinsic spin considered in the
previous sections, we can take
\begin{equation}
  \Pi = \sum_{m} \xi_{sm} \langle sm | \,,
\end{equation}
with the previous notations.  Then, with
\begin{equation}
  \Psi (p,q) = \Pi \rme^{-\frac{\rmi}{\hbar}\sum_i p_i \hat{q}_i } 
  \rme^{\frac{\rmi}{\hbar}\sum_i q_i \hat{p}_i} |\Psi\rangle \,,
\end{equation}
we obtain the prequantization
\begin{equation}
  \eqalign{
  \Gamma (\hat p_i) = -\rmi\hbar \frac{\partial}{\partial q_i} \,,\qquad
  \Gamma (\hat q_i) = q_i +\rmi\hbar \frac{\partial}{\partial p_i} \,,
  \\
  \Gamma (\hat J_i) = \hat{\mathcal{S}}_i -\rmi\hbar
  \Big( p_j\frac{\partial}{\partial p_k} - p_k\frac{\partial}{\partial
  p_j}\Big)  -\rmi\hbar
  \Big( q_j\frac{\partial}{\partial q_k} - q_k\frac{\partial}{\partial
  q_j}\Big) \, ,
  }
\end{equation}
which acts on vector-valued functions on ($p$-$q$) space.

\section{Irreducible representations and quantization}

A VCS representation will be irreducible if the intertwining operator
$\Pi$ is such that the only nonzero VCS wave functions are those of an
irrep.  Such irreps are found in VCS theory by a natural
generalization of the scalar coherent state construction.

It is known that a representation of a SGA $\mathfrak{g}$ extends
linearly to the complex extension $\mathfrak{g}^c$ of $\mathfrak{g}$.
The corresponding extension of a generic unitary representation $T$ of
the real group $G$ may not converge for all of $G^c$.  However, it may
be sufficient for the purpose of defining an irreducible coherent
state representation if the extension of $T$ is well-defined on
$\mathbb{H}$ for some subset $U(P)\subset P$ of a subgroup $P\subset
G^c$ which contains $H$. Let $\tilde M$ denote an irrep of $P\subset
G^c$ which restricts to a unitary irrep $M$ of $H\subset P$.  Now
suppose an intertwining operator can be found such that
\begin{equation}
  \psi(zg) =\Pi T(z)T(g) |\psi\rangle 
  = \tilde M(z) \psi(g) \, ,  
  \quad \forall\, z\in U(P) \,. 
  \label{eq:Pconstraint}
\end{equation}
We then say that the irrep $\Gamma$ is induced from the representation
$\tilde M$ of $P$.  It will be shown by examples in the following
sections that, for many categories of groups, there are natural
choices of $P$ and its representation $\tilde M$ for which the
corresponding VCS representation is irreducible.

Subgroups which lead to irreducible induced representations are
familiar in representation theory.  For example, if $G$ were
semisimple and the intrinsic symmetry group $H$ were a Levi subgroup,
a suitable subgroup $P\subset G^c$ would be the parabolic subgroup
generated by $H$ and the exponentials of a set of raising (or
lowering) operators.

Apart from imposing the stronger condition (\ref{eq:Pconstraint}), the
coherent state construction is the same as in
section~\ref{sect:VCSinduced}.  However, the stronger condition restricts
the set of coherent state wave functions to a subset with the result
that the coherent state representation becomes an irreducible
subrepresentation of that given in section~\ref{sect:VCSinduced}.

Now if a unitary coherent state representation $\Gamma$ of a
dynamical group $G$ induced from a representation $\tilde M$ of a
subgroup $P\subset G^c$ defines an irreducible representation of the Lie
algebra $\mathfrak{g}$ and if the representation $\tilde M$ satisfies
the equality
\begin{equation}
  \rmi\frac{\rmd}{\rmd t} \tilde M(\rme^{-\rmi At})\Big|_{t=0} =\tilde
  M(A)\equiv \hat\rho(A)\, , \quad A\in \mathfrak{p}\, ,
  \label{eq:QCondition}
\end{equation} 
then we say that $\Gamma$ is a quantization of the classical
representation of $\mathfrak{g}$ defined by $\hat\rho$.

Note, however, that for this quantization condition to be
satisfied, the classical representation corresponding to
the density $\hat \rho$ must define a representation $\tilde M$ of a
subalgebra $\mathfrak{p}\subset\mathfrak{g}^c$ that is contained in a
unique irrep of $\mathfrak{g}^c$ which restricts to a unitary irrep of
$\mathfrak{g}$.  This irrep of $\mathfrak{g}$ must integrate to a
(possibly projective) irrep of $G$.

The above VCS quantization of a classical model is a practical
expression of induced representation theory in the language of
geometric quantization.  Evidently the subgroup $P\subset G^c$ used to
construct an irreducible VCS induced representation defines an
invariant polarization of the tangent space at each point of the base
manifold $H\backslash G$ of the semi-classical bundle provided its
Lie algebra $\mathfrak{p}$ satisfies the conditions:
\begin{enumerate}
\item[(i)] $\hat\rho([A,B])=[\hat\rho(A),\hat\rho(B)]$ for any
  $A,B \in \mathfrak{p},$
\item[(ii)] ${\rm dim}_{\mathbb{R}}\, \mathfrak{g} +
  {\rm dim}_{\mathbb{R}}\, \mathfrak{h} = 2 
  \,{\rm dim}_{\mathbb{C}}\, \mathfrak{p},$
\item[(iii)] $\mathfrak{p}$ is invariant under the adjoint action 
  of $H$.
\end{enumerate}
The first condition ensures that the polarization is isotropic in the
sense that $\hat \Omega(A,B) =0$ for all $A,B\in \mathfrak{p}$.
The second condition ensures that $\mathfrak{p}$ is a maximal
subalgebra for which the first condition holds.  The final condition
ensures that the polarization is well-defined on $H \backslash G.$ In
all the examples we consider, these conditions are satisfied by the
Lie algebra $\mathfrak{p}\subset\mathfrak{g}^c$ used in the VCS
construction.

For the example of a particle with intrinsic spin considered in the
previous sections, we can take as a polarization the subalgebra
$\mathfrak{p}$ of $\mathfrak{g}^c$ spanned by the elements $\{ \hat I,
\hat J_i, \hat q_i\}$.  Let $\tilde M$ denote the representation of
$\mathfrak{p}$ which restricts to the previous representation $M$ of
$u(2)$ and to the zero representation of the abelian algebra spanned
by $\{ \hat q_i\}$; i.e., $\tilde M (\hat q_i)
= 0$.  Then, with $\Pi = \sum_{m} \xi_{sm} \langle sm | $ defined such
that
\begin{equation}
  \eqalign{
  \sum_{m} \xi_{sm} \langle sm | \hat J_i |\Psi\rangle = 
  \hat{\mathcal{S}}_i\sum_{m} \xi_{sm} \langle sm |\Psi\rangle \,, \\
  \sum_{m} \xi_{sm} \langle sm | \hat q_i |\Psi\rangle =0 \,,
  }
\end{equation}
so that $\langle sm|$ is a functional on a dense subspace of
$\mathbb{H}$, we obtain $p_i$-independent VCS wave functions and the
irreducible representation
\begin{equation}
  \eqalign{
  \Gamma (\hat p_i) = -\rmi\hbar \frac{\partial}{\partial q_i} \,,\qquad
  \Gamma (\hat q_i) = q_i  \,, \\
  \Gamma (\hat J_i) = \hat{\mathcal{S}}_i  -\rmi\hbar
  \Big( q_j\frac{\partial}{\partial q_k} - q_k\frac{\partial}{\partial
  q_j}\Big) \,,
  }
\end{equation}
of a full quantization.

\section{VCS inner products and Hilbert spaces}\label{subsec:VCSHilbert}

Let $U$ denote a Hilbert space with orthonormal basis $\{\xi_\nu\}$
for a finite-dimensional unitary irrep $M$ of a subgroup $H\subset G$.

We consider first the situation in which $U$ can be identified with an
$H$-invariant subspace of the Hilbert space $\mathbb{H}$ for some
unitary representation $T$ by an embedding $E :\, U\to\mathbb{H};\,
\xi_\nu \mapsto |\nu\rangle$.  The corresponding $\mathbb{H}\to U$
projection operator
\begin{equation}
  \Pi = \sum_\nu \xi_\nu \langle \nu |\,, 
  \label{eq:A.IntertwineOp}
\end{equation}
then satisfies the equation
\begin{equation}
   M(h)\Pi = \Pi T(h) \,,\quad \forall\ h\in H\,. 
  \label{eq:A.Hintertwining}
\end{equation}
Thus, $\Pi$ is an $H$-interwining operator and defines a set of VCS
wave functions
\begin{equation}
  \Psi(g) = \Pi T(g) |\Psi\rangle =\sum_\nu \xi_\nu
  \Psi_{\nu}(g)\,,\quad g\in G\,,\; |\Psi\rangle \in \mathbb{H} \,.
\end{equation}

Now, if $U$ is contained in a subrepresentation of $T$ which is a
direct sum of discrete series representations, the operator
\begin{equation}
  \mathbb{I} = \int_G \sum_\nu  
  T(g^{-1}) |\nu\rangle\langle\nu| T(g) \, \rmd v(g) \,,
\end{equation}
where $\rmd v$ is a left-invariant measure on $G$, is well-defined on
$\mathbb{H}$.  Moreover it commutes with the representation $T(g)$ of
any element $g\in G$.  Thus, by Schur's lemma, $\mathbb{I}$ acts as a
multiple of the identity on any irreducible subspace of $\mathbb{H}$.
Thus, an inner product is defined for the VCS wave functions by
\begin{equation}
  \eqalign{
  (\Psi, \Psi') = \langle \Psi|\mathbb{I}|\Psi'\rangle &= \displaystyle\int_G
  \Psi^*(g) \cdot \Psi'(g) \, \rmd v(g) \\ 
  &= \displaystyle\int_G\sum_\nu \Psi^*_\nu(g) \Psi'_\nu(g) \, \rmd
  v(g) \,.
  }
  \label{eq:VCSIP}
\end{equation}
However, because $\Psi(hg) = M(h) \Psi(g)$ for $h\in H$, the scalar
product in $U$ satisfies
\begin{equation}
  \Psi^*(hg)\cdot\Psi'(hg) = \Psi^*(g)\cdot\Psi'(g) 
\end{equation}
and the integral over $G$ in equation~(\ref{eq:VCSIP}) can be restricted to
an integral over the coset space $H\backslash G$ with respect to the
left $H$-invariant measure inherited from $G$.

The above construction works when $M$ is a subrepresentation of the
restriction of $T$ to $H\subset G$.  If $M$ is not a subrepresentation
but is contained in a direct integral decomposition of the restriction
of $T$ to $H$, then it is still possible to define an
$H$-intertwining operator by equation~(\ref{eq:A.IntertwineOp}) that
satisfies equation~(\ref{eq:A.Hintertwining}) albeit with $\{ \langle
\nu|\}$ defined as a set of functionals on a dense subspace
$\mathbb{H}_D$ of $\mathbb{H}$.  It can then happen that the integral
expression for $\mathbb{I}$ may not converge.  However, the
corresponding integral over $H\backslash G$ may converge and, if so,
it defines an inner product for VCS wave functions in parallel with
Mackey's construction of inner products for induced representations.
Inner products for more general VCS representations are constructed by
K-matrix methods \cite{rowe88}.

The Hilbert space of all VCS wave functions that satisfy the
constraint equation (\ref{eq:LeftAction}) and are normalizable with
respect to the above-defined inner product is that of the standard
representation of $G$ induced from the representation $M$ of the
subgroup $H\subset G$.  The subspace of VCS wave functions that
satisfy the stronger constraint condition (\ref{eq:Pconstraint})
for a suitable polarization is the Hilbert space for an irreducible
induced representation. 

\section{Examples of VCS representations} 

The $SU(3)$ and rigid rotor models provide insightful and
representative examples of the VCS quantization methods.  Despite its
apparent simplicity, the quantization of rotational models is
considerably more difficult than traditional canonical problems with
three degrees of freedom.  The difficulties arise from the nontrivial
geometry of the phase spaces and the possibility of intrinsic degrees
of freedom.  However, the VCS quantization techniques handle these
problems with ease.  In the following, algebraic formulations of both
the $SU(3)$ and rotor models will be given, and the techniques of the
previous sections will be used to investigate their classical,
semi-classical, and quantal realizations with intrinsic degrees of
freedom.

\subsection{Coherent state representations of SU(3)}\label{sec:SU(3)}

An $su(3)$ model was first formulated as an algebraic model of nuclear
rotations by Elliott~\cite{elliott}.  It was followed by the $su(3)$
quark model of Gell-Mann and Ne'eman~\cite{gellmann}.  These models
have enjoyed enormous successes partly because of their simplicity;
the $su(3)$ algebra is semi-simple and has a straightforward and well
understood representation theory; it is also compact and its unitary
irreps are finite dimensional.  VCS theory was applied to $su(3)$
in~\cite{RLeBR} and reviewed in~\cite{rowe96,rowe98}.

Let $\{ C_{ij}; i,j=1,2,3 \}$ be the standard basis for
$gl(3,\mathbb{C}) \simeq u(3)^c $ with commutation relations
\begin{equation}
  \label{eq:LieAlgebragl(3,C)}
  [C_{ij},C_{kl}] = \delta_{jk}C_{il} - \delta_{il} C_{kj} \, .
\end{equation}
Then $su(3)$ is the real linear span of the hermitian combinations
\begin{equation}
  \label{eq:su(3)CartesianBasis}
  \eqalign{
  J_{ij} = -\rmi (C_{ij} - C_{ji})\, , & i< j \, , \\
  Q_{ij} = (C_{ij}+C_{ji})\, , & i< j\, ,\\
  H_i = (C_{ii} - C_{i+1,i+1})\, , \quad& 1\leq i \leq 2\, .
  }
\end{equation}

Let $T$ denote the regular representation of the group $SU(3)$.  It
can be extended to a representation of $SL(3,\mathbb{C})$ on the
algebraic direct sum of the irreps of $SU(3)$, which is dense in the
regular representation.  As usual we denote by $A \to \hat A = T(A)$
the corresponding representation of the Lie algebra
$sl(3,\mathbb{C})$.  The coherent state methods outlined lead to
several classes of $su(3)$ representations corresponding to: classical
representations, semi-classical representations of a partial
quantization, the induced representations of prequantization, and the
irreducible unitary representations of a full quantization.

\subsubsection{Classical representations}

Scalar coherent state techniques lead to a classical representation as
follows.  Let $|0\rangle$ be some state in the Hilbert space
$\mathbb{H}$ of the representation $T$ for which
\begin{equation}
  \eqalign{
  \langle 0| \hat C_{ij} |0\rangle = 0 \,, \quad i\not= j \,,\\
  \langle 0| \hat H_i |0\rangle = \nu_i \,.
  }
\end{equation}
By the standard moment map, a classical density $\rho \in su(3)^*$ is
defined by
\begin{equation}
  \label{eq:RhoSU(3)}
  \rho( X ) = \langle 0 | \hat X | 0
  \rangle\, , \quad X \in su(3) \,,
\end{equation}
and extended linearly to elements of $su(3)^c$ in the usual way by
setting $\rho (X+\rmi Y) = \rho(X) + \rmi \rho(Y)$.  A classical phase space
is defined as the coadjoint orbit $\mathcal{O}_\rho = \{ \rho_g; g \in
SU(3) \},$ where
\begin{equation}
  \label{eq:DefSU(3)Rhog}
  \rho_g(A) =  \rho (A(g)) = \langle 0 |T(g)\hat A
  T(g^{-1})|  0 \rangle \, ,  \quad A \in su(3) \, .  
\end{equation}
This phase space is diffeomorphic to the factor space
$H_\rho\backslash SU(3)$, where $H_\rho$ is the isotropy subgroup
\begin{equation}
  H_\rho = \{ h\in SU(3) \,|\; \rho_h = \rho \, \} \,.
\end{equation}
We consider the generic situation, in which $H_\rho$ is the Cartan
subgroup with Lie algebra spanned by $H_1$ and $H_2$.  (When $\nu_1$
or $\nu_2 $ is zero, for example, $H_\rho$ is a larger subgroup and
the construction simplifies.)  A classical representation of
$su(3)$ is then defined in which an element
$A \in su(3)^c $ maps to a function $\mathcal{A}$ on
$H_\rho\backslash SU(3)$ with values
\begin{equation}
  \mathcal{A}(g) =  \rho_g (A) = \rho (A (g)) \,.
\end{equation}
The Poisson bracket for this classical representation is defined in
the standard way by
\begin{equation}
  \label{eq:SU(3)ClassicalPoissonBracket}
  \{ \mathcal{A}, \mathcal{B} \}(g) = \omega_g(A,B) = 
  - \frac{\rmi}{\hbar} \rho_g([A,B]) \, ,
\end{equation}
for $A,B \in su(3)$.

The above representation can be obtained in explicit form in terms of
suitable coordinate charts for $H_\rho\backslash SU(3)$ (see examples
in~\cite{scalar}).  For example, Murnaghan~\cite{Murnaghan} has shown
that an $SU(3)$ matrix can be parameterized by the factorization
\begin{equation}
  \label{eq:SU3parameters}
  g(\xi,\alpha,\beta) = \rme^{-\rmi(\xi_1H_1+\xi_2H_2)}
  g_{23}(\alpha_1,\beta_1)\, g_{13}(\alpha_2,\beta_2)\, 
  g_{12}(\alpha_3,\beta_3)  \,, 
\end{equation}
where 
\begin{equation}
  g_{23}(\alpha,\beta) = \left(\begin{array}{ccc} 1&0&0\\
  0& \cos\beta & - \rme^{-\rmi\alpha}\sin\beta  \\
  0& \rme^{\rmi\alpha}\sin\beta & \cos\beta  \end{array} \right)
\end{equation}
and $g_{13}$ and $g_{12}$ are similarly defined.  Since the first
factor on the rhs of equation~(\ref{eq:SU3parameters}) is an element
of the isotropy subgroup $H_\rho$, this parameterization leads to a
classical representation of the $su(3)$ algebra in terms of functions
of the $(\alpha,\beta)$ coordinates.

Now observe that the first two factors on the rhs of
equation~(\ref{eq:SU3parameters}) are elements of a $U(2)\subset
SU(3)$ subgroup.  This suggests a fibration of the classical phase
space $H_\rho\backslash SU(3)$ as an intrinsic $H_\rho\backslash U(2)$
phase space over an extrinsic $U(2)\backslash SU(3)$ phase space.
Because the representation theory of $U(2)$ is well known, this
greatly facilitates the quantization process.

\subsubsection{Semi-classical representations} 

The intrinsic symmetry algebra $u(2)$ suggested by the above
parameterization of $SU(3)$ is spanned by  $H_1$ and the elements of an
$su(2)$ algebra 
\begin{equation}
  S_z = \case12 H_2 \,, \quad 
  S_x = \case12 ( C_{23}+C_{32}) \,,
  \quad S_y = -\case12 \rmi( C_{23}-C_{32}) \,.
\end{equation}
Thus, for the intrinsic degrees of freedom of the $SU(3)$ classical
phase space to be quantizable, it is required that $\nu_2$ should be
an integer.  Moreover, in order that it should be an $su(2)$ highest
weight and uniquely define an $su(2)$ irrep, it should be a positive
integer.  The representation label $\nu_1$ is not so constrained.
For, if it is not an integer, the only consequence is that the
associated representation of the $u(1)$ algebra integrates to a
unitary projective representation of U(1), i.e., a unitary
representations of a covering group of $U(1)$.  This is not possible
for $su(2)$ because the group $SU(2)$ is simply connected; it is its
own universal covering group.

Let $M$ denote a unitary (possibly projective) irrep of the $U(2)$
intrinsic symmetry group of highest weight $(\nu_1,\mu)$ (with $\mu$ a
positive integer) on a Hilbert space $U$.  Let $E : U\to \mathbb{H}$
be an embedding of $U$ in the regular representation $\mathbb{H}$ and
let $\Pi : \mathbb{H}\to U$ be the corresponding orthogonal projection
with respect to the inner product for $\mathbb{H}$.  The embedding $E$
is required to be such that
\begin{equation}
  \Pi \hat A E = M(A)\,,\quad \forall\, A\in u(2)\subset su(3) \,.
\end{equation}
Now define
\begin{equation}
  \hat\rho (A) = \Pi \hat A E \,, \quad A\in su(3) \,,
\end{equation}
and assume, for convenience, that $E$ is chosen such that
\begin{equation}
  \label{eq:ChoiceOfU(2)Embedding}
  \hat\rho(C_{12}) = \hat\rho(C_{13}) = \hat\rho(C_{21}) =
  \hat\rho(C_{31}) = 0 \,.
\end{equation}
Then
\begin{equation}
  \hat\rho (H_1 + \case12  H_2) = 
   \nu_1+\case12 \mu \,, \quad\hat\rho (S_i) =\hat S_i \,,
\end{equation}
where we draw attention to the fact $(H_1 + \case12  H_2)$ commutes
with the $su(2)$ operators $\{ S_i\}$ but $H_1$ on its own does not.

A partial quantization of $su(3)$ is now defined as a semi-classical
representation in which an element $A\in su(3)$ is mapped to a
$U(2)$-equivariant operator-valued function $\hat{\mathcal{A}}$ on
$SU(3)$ with values
\begin{equation}
  \hat{\mathcal{A}}(g) = \hat{\rho}(A(g)) \,.
\end{equation}

Note that, because
\begin{equation}
  \hat{\mathcal{A}}(hg) = M(h) \hat{\mathcal{A}}(g) M(h^{-1}) \,, \quad 
  \forall\ h \in U(2) \, ,
\end{equation}
it is sufficient to evaluate the classical operator-valued functions
and their Poisson brackets on a set of $U(2)\backslash SU(3)$ coset
representatives (cf.\ section~\ref{sec:Partial}).  Thus, making use of
the Murnaghan factorization of equation~(\ref{eq:SU3parameters}), a
semi-classical representation is defined over a set of
$U(2)\backslash SU(3)$ coset representatives $K = \{
k(\alpha,\beta)\}$ with
\begin{equation}
  k(\alpha,\beta) =  g_{13}(\alpha_2,\beta_2)\, g_{12}(\alpha_3,\beta_3) 
  \,,  \label{eq:Mcosetreps}
\end{equation}
for a suitable range of $(\alpha,\beta)$ values.  

The expressions for this semi-classical representation as
operator-valued functions of $(\alpha,\beta)$ can be worked out.
However, they are expressed more simply in terms of coset
representatives
\begin{equation}
  K = \{  \rme^{Y(y)} \rme^{Z(z)}\} \,, \label{eq:YZcosetreps}
\end{equation}
for $H^c\backslash G^c$ for which $Y(y)$ and $Z(z)$ are linear
combinations of commuting Lie algebra elements, i.e.,
\begin{equation}
  Y(y) = y_2C_{21}+y_3C_{31} \,,\quad Z(z)=z_2C_{12} + z_3C_{13} \,.
\end{equation}
From the identities
\begin{equation}
  \fl\eqalign{
  e^{Y(y)} e^{Z(z)} C_{12} e^{-Z(z)}e^{-Y(y)} &=
  C_{12} -y_2 H_1 + y_3 C_{32} - y_2^2 C_{21} - y_2 y_3 C_{31} \, ,
  \\
  e^{Y(y)} e^{Z(z)} C_{13} e^{-Z(z)}e^{-Y(y)} &=
  C_{13} -y_3 (H_1 + H_2) + y_2 C_{23} - y_2 y_3 C_{21} - y_3^2 C_{31}
  \, , \\
  e^{Y(y)} e^{Z(z)} C_{23} e^{-Z(z)}e^{-Y(y)} &=
  (1+y_2 z_2)C_{23} -y_3 z_2(H_1 + H_2) \\
  &\quad - y_3(1+y_2z_2) C_{21} + z_2
  C_{13} - z_2 y_3^2 C_{31} \,,
  }
\end{equation}
it follows that the semi-classical representations of the elements
$C_{12}$, $C_{13}$ and $C_{23}$ are given by
\begin{equation}
  \eqalign{
  \hat{\mathcal{C}}_{12}(y,z) &= y_3 \hat S_- -  y_2
  (\nu_1+\case12 \mu)\hat I + y_2\hat S_z \,, \nonumber \\
  \hat{\mathcal{C}}_{13}(y,z) &= y_2 \hat S_+ - y_3
  (\nu_1+\case12 \mu) \hat I -  y_3 \hat S_z \,, \nonumber \\
  \hat{\mathcal{C}}_{23}(y,z) &= (1+y_2 z_2) \hat S_+ - y_3 z_2
  (\nu_1+\case12 \mu)\hat I - y_3 z_2 \hat S_z \, ,
  }
\end{equation}
where $\hat S_\pm = M(S_x \pm \rmi S_y)$ and $\hat I$ is the unit operator
on $U$.  The representations of all elements of $su(3)$ can be derived
in this fashion. Calculating the semi-classical Poisson bracket $\{
\hat{\mathcal{C}}_{12}, \hat{\mathcal{C}}_{23} \}(y,z)$ as defined by
equation~(\ref{eq:SemiclassPoissonBracketDecomposition}), we find that
the quantal part is given by
\begin{equation}
  \label{eq:PBQuantalPart}
  [\hat{\mathcal{C}}_{12}(y,z),\hat{\mathcal{C}}_{23}(y,z)] = 
  y_2(1+y_2z_2)\hat S_+ -y_3^2 z_2 \hat S_-  -2y_3(1+y_2 z_2)\hat S_z
  \, , 
\end{equation}
and the classical part by
\begin{equation}
  \label{eq:PBClassicalPart}
  \fl \rmi \hbar
  \hat\Omega(X_{\hat{\mathcal{C}}_{12}},X_{\hat{\mathcal{C}}_{23}})
   (y,z) = - y^2_2z_2\hat S_+ +  y_3^2 z_2 \hat S_-
  - y_3 (\nu_1+\case12 \mu) \hat I  + y_3(1+2y_2z_2) \hat S_z \, . 
\end{equation}
Together, these components give
\begin{equation}
  \label{eq:PBBothParts}
  \eqalign{
  \rmi\hbar \{ \hat{\mathcal{C}}_{12}, \hat{\mathcal{C}}_{23} \}(y,z) &=
  [\hat{\mathcal{C}}_{12}(y,z),\hat{\mathcal{C}}_{23}(y,z)] + \rmi \hbar
  \hat\Omega( X_{\hat{\mathcal{C}}_{12}},
  X_{\hat{\mathcal{C}}_{23}})  (y,z)   \\
  &= y_2 \hat S_+ - (\nu_1+\case12 \mu)y_3  \hat I -  y_3
  \hat S_z \,, \\
  &= \hat{\mathcal{C}}_{13}(y,z) \, ,
  }
\end{equation}
as required for a semi-classical representation. However, as we now
show, the representations of prequantization and their commutation
relations are easier to derive, and those of the irreducible
representations of a full quantization are even simpler.

\subsubsection{The induced representations of prequantization} 

To be quantizable, the irrep $M$ of the $u(2)\subset su(3)$ subalgebra
of a semi-classical representation should be a $u(2)$ irrep contained
in some unitary representation $T$ of $su(3)$.  This condition
requires that $\nu_1$ also be a positive integer.  Thus, we now
suppose that $M$ is an irrep of $u(2)$ on an intrinsic Hilbert space
$U$ with highest weight $(\lambda,\mu)$, where $\lambda$ and $\mu$ are
both positive integers.  This representation extends to a
representation of the $U(2)$ group.

Let $T$ be an abstract representation of $SU(3)$ on a Hilbert space
$\mathbb{H}$ and suppose the irrep $M$ of $U(2)$ is contained in $T$.
Then there exists a $U(2)$-intertwining operator $\Pi :\mathbb{H}\to
U$ satisfying
\begin{equation}
  \label{eq:su(3)IntertwiningRelation}
  \Pi T(h) = M(h) \Pi\, , \quad \forall\ h \in U(2)\, .
\end{equation}
For example, suppose $V\subset \mathbb{H}$ is an irreducible
$U(2)$-invariant subspace of $\mathbb{H}$ with orthonormal basis $\{
|sm\rangle; m = -s,\ldots, +s,\, s= \mu/2\}$ and the intertwining
operator
\begin{equation}
  \Pi = \sum_m \xi_{sm} \langle sm| 
\end{equation}
maps this basis to a corresponding basis $\{ \xi_{sm}\}$ for $U$.

The VCS wave functions are now defined, over the coset representatives
of equation~(\ref{eq:YZcosetreps});
\begin{equation}
  \Psi(y,z) = \Pi \rme^{\hat Y(y)} \rme^{\hat Z(z)} |\Psi\rangle \, ,
\end{equation}
with $\hat Y(y) = y_2\hat C_{21}+y_3\hat C_{31} $ and $\hat
Z(z)=z_2\hat C_{12} + z_3\hat C_{13}$.  Thus, for example, the
representation $\Gamma(C_{12})$ of the element $C_{12}\in su(3)^c$ is
given immediately by
\begin{equation}
  [\Gamma(C_{12})\Psi](y,z) = \Pi  \rme^{\hat Y(y)} \rme^{\hat Z(z)}
  \hat C_{12} |\Psi\rangle = \frac{\partial}{\partial z_2} \Psi(y,z) \,.
\end{equation}
The representations of other $su(3)^c$ elements are obtained almost as
easily.  For example, the expression for one of the most complicated
elements, defined by
\begin{equation}
  [\Gamma(C_{21})\Psi](y,z) = \Pi  \rme^{\hat Y(y)} \rme^{\hat Z(z)}
  \hat C_{21} |\Psi\rangle \,,
\end{equation}
is obtained from the identities
\begin{equation}
  \eqalign{
  \rme^{\hat Z(z)}\hat C_{21}
  =\Big(\hat C_{21} +z_2(\hat C_{11}-\hat C_{22}) -z_3 \hat{C}_{23}  
  - z_2^2 \hat C_{12} -  z_2z_3\hat C_{13}\Big) \rme^{\hat Z(z)} \,, \\
  \rme^{\hat Y(y)}(\hat C_{11}-\hat C_{22})
  = (\hat C_{11}-\hat C_{22} + 2y_2\hat C_{21} + y_3 \hat C_{31} )
  \rme^{\hat Y(y)}\,,\\
  \rme^{\hat Y(y)}\hat C_{23}
  = (\hat C_{23}  -y_3\hat C_{21} ) \rme^{\hat Y(y)}\,.
  }
\end{equation}
It follows that
\begin{eqnarray}
  \fl \Gamma(C_{21})
  = (1+y_3z_3) \frac{\partial}{\partial y_2} -z_2\hat S_z -z_3 \hat
  S_+  \nonumber \\ \lo +z_2 \Big((\lambda+\case12 \mu)
  + 2y_2\frac{\partial}{\partial y_2} + y_3\frac{\partial}{\partial y_3}
  -z_2 \frac{\partial}{\partial z_2} -z_3 \frac{\partial}{\partial
  z_3}\Big)  \,. 
\end{eqnarray} 
Similarly, one obtains
\begin{equation}
  \Gamma(H_1) =  (\lambda+\case12 \mu) +
  2y_2\frac{\partial}{\partial y_2} +
  y_3\frac{\partial}{\partial y_3} -2z_2\frac{\partial}{\partial z_2}
  -z_3\frac{\partial}{\partial z_3} 
  - \hat S_z \,.
\end{equation}
It is readily checked that these operators satisfy the commutation
relations
\begin{equation}
  \eqalign{
  [\Gamma(C_{12}),\Gamma(C_{21})] = \Gamma(H_1) \,, \\
  {[}\Gamma(H_1),\Gamma(C_{12})] = 2\Gamma(C_{12}) \,, \\
  {[}\Gamma(H_1),\Gamma(C_{21})] = -2\Gamma(C_{21}) \,.
  }
\end{equation}

\subsubsection{The irreducible representations of a full quantization}

For an induced VCS representation of $su(3)$ to be irreducible, the
map $\Pi : \mathbb{H}\to U$ must be chosen such that it intertwines a
representation of a larger subgroup $P\subset SU(3)^c$ corresponding
to a polarization.  Since an irrep of $SU(3)$ is uniquely defined by
its highest weight $(\lambda,\mu)$, it is also uniquely defined by an
irrep $\tilde M$ of the $\mathfrak{p} \subset su(3)^c$ subalgebra
spanned by the elements $\{ C_{23}, C_{32}, H_1,H_2\}$ of the $u(2)$
subalgebra, considered for prequantization, together with the
operators $\{C_{21},C_{31}\}$.  The appropriate irrep is then one for
which
\begin{equation}
  \tilde M(A) = M(A) \,, \quad \forall\, A\in u(2) \, ,
\end{equation}
and
\begin{equation}
 \tilde M(C_{21}) = \tilde M(C_{31}) = 0 \,.
\end{equation}
Thus, we take for $P$ the parabolic subgroup of $SU(3)^c$ generated by
exponentiating the Lie algebra $\mathfrak{p}$.  The representation
$\tilde M$ of $\mathfrak{p}$ is likewise exponentiated to an irrep of
$P$.  Now if $\Pi$ is an intertwining operator such that
\begin{equation}
  \Pi T(p) = \tilde M(p) \Pi\,, \quad\forall\ p \in P \,,
\end{equation}
then VCS states are defined by
\begin{equation}
  \Psi(z) = \Pi \rme^{\hat{Z}} |\Psi\rangle \,,
\end{equation}
with $\hat{Z}(z)=z_2\hat C_{12} + z_3\hat C_{13}$.  It is immediately
seen that such wave functions are the $y$-independent subset of those
of the prequantization of the previous section. Thus, one immediately
obtains the operators of an irrep with, for example,
\begin{equation}
  \eqalign{
  \Gamma(C_{12}) =  \frac{\partial}{\partial z_2} \,,\\
  \Gamma(C_{21})  =  z_2 \Big( (\lambda+\case12 \mu) - 
  z_2 \frac{\partial}{\partial z_2}
  -z_3 \frac{\partial}{\partial z_3}\Big)
  - z_2 \hat S_z -z_3 \hat S_+  \,, \\
  \Gamma(H_1) = (\lambda+\case12 \mu) -2z_2\frac{\partial}{\partial z_2}
  -z_3\frac{\partial}{\partial z_3} - \hat S_z \,.
  }
\end{equation}
This is a standard holomorphic induced representation. 

An inner product for this representation is defined such that the
representation of the real $su(3)$ algebra is by Hermitian operators.
This inner product leads to an explicit construction of an orthonormal
basis for an irrep~\cite{RR97}.

\subsubsection{The relationship between VCS and scalar coherent state
representations}

A VCS representation can also be expressed as a scalar coherent state
representation.  However, contrary to what one might expect, the
latter is generally more complicated.  Consider the above example of a
VCS representation of $SU(3)$.  An equivalent scalar coherent state
representation is given by realizing the vectors $\{ \xi_{sm}\}$ in a
coherent state representation for $U(2)$ for which $\xi_{sm}$, with
$s=\mu/2$, becomes a real function of $SO(2)$:
\begin{equation} 
  \xi_{sm} (\theta) = \langle \lambda\mu | \rme^{\rmi\theta \hat S_y} |sm
  \rangle\, .
\end{equation}
A holomorphic VCS wave function is then expressed as a scalar
coherent state function by observing that
\begin{equation}
  \eqalign{
  \Psi (\theta,z) &= \sum_m \xi_{sm} (\theta) \langle sm|
  \rme^{\hat Z(z)} |\Psi\rangle \nonumber \\
  &= \langle \lambda \mu | \rme^{\rmi\theta \hat S_y} 
  \Bigl( \sum_m |sm \rangle
  \langle sm | \Bigr) \rme^{\hat Z(z)} | \Psi \rangle \nonumber\\
  &= \langle \lambda\mu |\rme^{\rmi\theta \hat S_y}\rme^{\hat Z(z)} 
  |\Psi\rangle\, .
  }
  \label{eq:VCSasScalarFunction}
\end{equation}

The advantage of the VCS representation is that it subsumes all the
properties of the chosen subgroup, in this case $U(2)$, and thereby
avoids having to reproduce them in the expression of the larger group,
in this case $SU(3)$.  However, it is useful to know that a VCS
representation can always be expressed as a scalar coherent state
representation because it means that any results proved for a scalar
CS representation automatically apply, with appropriate
interpretation, to a VCS representation.

\subsection{Rigid rotor models} 

A classical rigid rotor is characterized by a rigid intrinsic
structure.  Thus, the dynamical variables of a rigid rotor are its
orientation and angular momentum.  We consider here an algebraic rotor
model with an algebra of observables spanned by the components of the
angular momentum and the moments of the inertia tensor for the rotor.

The moments $\{I_{ij}\}$ of the inertia tensor (in a Cartesian basis) can
be viewed as the elements of a real symmetric $3\times 3$ matrix.  Given
values for these observables, the orientation of a rotor is defined
(with some ambiguity) by the rotation $\Omega\in SO(3)$ that brings the
inertia tensor to diagonal form,
\begin{equation}
  \label{eq:DefinitionInertiaTensor}
  \overline{I}_{ij} = \Omega I  \Omega^{-1}_{ij} = 
  \delta_{ij} {I}_i\, , 
\end{equation}
where $({I}_1,{I}_2,{I}_3)$ are fixed
intrinsic moments of inertia.

Because the inertia tensor is a function only of orientation, its
components commute,
\begin{equation}
  \label{eq:InertiaTensorCommutes}
  [I_{ij},I_{kl}] = 0 \, ,
\end{equation}
and span an algebra isomorphic to $\mathbb{R}^6$.  The angular
momentum $L$ has Cartesian components $\{ L_i; i=1,2,3 \}$
which span an $so(3)$ Lie algebra,
\begin{equation}
  \label{eq:su(3)LieAlgebra}
  [L_i,L_j] = \rmi \hbar L_k \, , \quad i,j,k\ {\rm cyclic}.
\end{equation}
The inertia tensor is defined, by (\ref{eq:DefinitionInertiaTensor}),
to be a rank-2 Cartesian tensor.  Thus, it obeys the commutation
relations
\begin{equation}
  \label{eq:ImIsRank2SphericalTensor}
  [I_{ij},L_k] = \rmi \hbar \sum_l (\varepsilon_{lik} I_{lj} + 
  \varepsilon_{ljk} I_{li} ) \, .
\end{equation}
Together, the moments of inertia and the angular momenta span a SGA
for the rotor that is isomorphic to the semidirect sum algebra
$[\mathbb{R}^6]so(3)$ with $\mathbb{R}^6$ as its ideal.  This algebra
is known as the \emph{rotor model algebra} (RMA).

The corresponding dynamical group obtained by exponentiating the RMA
is the \emph{rotor model group} (RMG), a group isomorphic to the
semidirect product $[\mathbb{R}^6]SO(3)$.  An element of the RMG is a
pair $(Q,\Omega)$, with $Q \in \mathbb{R}^6$ and $\Omega \in
SO(3)$ and the group product is given by 
\begin{equation}
  \label{eq:RMGgroupmultiplicationrule}
  (Q_1, \Omega_1)\circ (Q_2, \Omega_2) = (Q_1 + \Omega_1 Q_2
  \Omega_1^{-1}, \Omega_1 \Omega_2) \, .
\end{equation}

This group and its Lie algebra have many classical and quantal
representations.  The classical representations of rigid rotor models
and Euler's equations for their Hamiltonian dynamics are well known.
The quantization of the rigid rotor was given by
Casimir~\cite{casimir} and is well known in nuclear~\cite{bohr} and
molecular physics (cf.\ ref.~\cite{townes} for a review).  The route
from classical representations of the rotor to the unitary
representations of quantum mechanics is an illuminating example for
both the methods of induced representations and of geometric
quantization. We show here that the classical and quantal
representations have simple expressions in coherent state and VCS
theory.

\subsubsection{Classical representations} 

A classical representation of a rigid rotor can be derived from any
abstract unitary representation $T$ of the RMA $[\mathbb{R}^6]so(3)$
on a Hilbert space $\mathbb{H}$.  Let $\hat A = T(A)$ for $A\in
[\mathbb{R}^6]so(3)$.  Let $|0\rangle$ be a normalized state in
$\mathbb{H}$ and $\rho_0$ a corresponding density satisfying
\begin{equation}
  \label{eq:RMADensityOrigin}
  \rho_0(L_i) = \langle 0 | \hat{L}_i |0\rangle =  0 \, , \qquad
  \rho_0(I_{ij}) = \langle 0 | \hat{I}_{ij} | 0\rangle =
  \overline{\Im}_{ij} = \delta_{ij} {\Im}_i \, ,
\end{equation}
with $i,j = 1,2,3$ and ${\Im}_i \in \mathbb{R}$.  Then $\rho_0$ is the
element of the dual ${\rm RMA}^*$ that represents a classical state
with zero angular momentum and orientation such that the inertia
tensor $\overline\Im$ is diagonal, i.e., the principal axes of this
inertia tensor coincide with those of the space-fixed coordinate
frame.  As usual, many classical irreps (in this case with different
principal moments of inertia $\{\Im_i\}$) can be derived from a given
unitary representation $T$ by different choices of $\rho_0$.

Starting with a density $\rho_0$, a classical phase space for the
rotor is the coadjoint orbit
\begin{equation}
  \label{eq:ZeroIntrinsicCoadjointOrbit}
  \mathcal{O}_\rho = \{ \rho_{(Q,\Omega)};\, (Q,\Omega) \in
  [\mathbb{R}^6]SO(3) \} 
\end{equation}
of the RMG in ${\rm RMA}^*$, where $\rho_{(Q,\Omega)}$ is defined by
\begin{equation}
  \label{eq:RMADensities}
  \eqalign{
  \rho_{(Q,\Omega)}(L_i) &= \langle 0 | T(Q,\Omega) \hat{L}_i
  T((Q,\Omega)^{-1})|0\rangle \, , \nonumber \\
  \rho_{(Q,\Omega)}(I_{ij}) &= \langle 0 | T(Q,\Omega)\hat{I}_{ij}
  T((Q,\Omega)^{-1})| 0\rangle \, .
  }
\end{equation}
The set of functions $\{ \Im_{ij}, \mathcal{L}_i; \; i,j = 1,2,3\}$,
defined by
\begin{equation}
  \label{eq:ClassicalRMA}
  \eqalign{
  \Im_{ij}(Q,\Omega) = \rho_{(Q,\Omega)} (I_{ij})= \sum_k \Im_k
  \Omega_{ki} \Omega_{kj} \, , \\
  \mathcal{L}_l(Q,\Omega) = \rho_{(Q,\Omega)}(L_l) 
  =-\hbar \sum_{ijk} \varepsilon_{ijk} Q_{ij} (\Im_i-\Im_j)\Omega_{kl}\,,
  }
\end{equation}
are then a basis for a classical representation of the RMA with
Poisson brackets
\begin{equation}
  \eqalign{
  \{ \Im_{ij}, \Im_{kl} \}(Q,\Omega) &= -\frac{\rmi}{\hbar}
  \rho_{(Q,\Omega)}([I_{ij},I_{kl}]) = 0 \, , \\
  \{ \mathcal{L}_i, \mathcal{L}_j \}(Q,\Omega) &= -\frac{\rmi}{\hbar}
  \rho_{(Q,\Omega)}([L_i,L_j]) = \sum_k \varepsilon_{ijk}
  \mathcal{L}_k(Q,\Omega)\, , \\
  \{ \Im_{ij}, \mathcal{L}_k \}(Q,\Omega) &= -\frac{\rmi}{\hbar}
  \rho_{(Q,\Omega)}([I_{ij},L_k]) \\ 
  &= \sum_l (\varepsilon_{lik} \Im_{lj}(Q,\Omega) +
  \varepsilon_{ljk} \Im_{li}(Q,\Omega) ) \,.
  }
\end{equation}

If the three principal moments of inertia
$\{{\Im}_1,{\Im}_2,{\Im}_3\}$, are all different, then the subgroup of
rotations that leave the density $\rho_0$ invariant under the
coadjoint action is the discrete group $D_2$ generated by rotations
through angle $\pi$ about the principal axes and the isotropy subgroup
of the phase space is the semidirect product $[\mathbb{R}^3]D_2$,
where $\mathbb{R}^3 \subset \mathbb{R}^6$ is the subgroup generated by
the diagonal moments $\{ I_{ii}, i=1,2,3 \}$. The phase space
$\mathcal{O}_0 \simeq [\mathbb{R}^3]D_2 \backslash
[\mathbb{R}^6]SO(3)$ is then symplectomorphic to the cotangent bundle
$T^* (D_2\backslash SO(3))$. This orbit is the phase space of an
\emph{asymmetric top}. If two of the principal moments of inertia are
equal, e.g., $\Im_1=\Im_2 \neq \Im_3$, then the subgroup of rotations
that leave $\rho_0$ invariant is $D_{\infty}$, a group comprising
rotations about the symmetry axis and rotations through angle $\pi$
about perpendicular axes.  The isotropy subgroup of the phase space is
then $[\mathbb{R}^4]D_\infty$, where $\mathbb{R}^4\subset
\mathbb{R}^6$ is the subgroup generated by $\{ I_{ii}, i=1,2,3 \}$ and
$I_{12}$.  The phase space $\mathcal{O}_0 \simeq
[\mathbb{R}^4]D_\infty \backslash [\mathbb{R}^6]SO(3)$ is then
symplectomorphic to the cotangent bundle $T^* (D_{\infty}\backslash
SO(3))$ which is the phase space of a \emph{symmetric top}.

The phase space for a symmetric top is of lower dimension than that of
an asymmetric top.  One of the reasons for this difference is that
there is no element of the RMA that can generate a boost in the
component of the angular momentum about a symmetry axis.  Thus, when
$\Im_1=\Im_2$, the component of the angular momentum along the 3-axis
is a constant of the motion with value given by that at $\rho_0$.
This condition does not mean that a symmetric top cannot rotate about
its symmetry axis.  It means only that it rotates about its symmetry
axis with a constant angular momentum.  Thus, the component of angular
momentum along the symmetry axis of a symmetric top is appropriately
regarded as an intrinsic (gauge) degree of freedom.

Consider, for example, a symmetric top representation for which $\Im_1
= \Im_2 \neq \Im_3$ and, instead of $|0\rangle$, consider a normalized
state $|K\rangle$ and corresponding density $\rho_0^{(K)}$ for which
\begin{equation}
  \label{eq:RMAModifiedDensityOrigin}
  \eqalign{
  \rho_0^{(K)}(L_i) &= \langle K | \hat{L}_i |K\rangle = 0 \, , \quad
  i=1,2\, , \\ 
  \rho_0^{(K)}(L_3) &= \langle K | \hat{L}_3 |K\rangle = K \, , \\
  \rho_0^{(K)}(I_{ij}) &= \langle K | \hat{I}_{ij} |K\rangle = \delta_{ij}
  {\Im}_i \,,
  }
\end{equation}
where $K$ is a real constant.  The density $\rho_0^{(K)}\in
{\rm RMA}^*$ is that of a symmetric top with its axis of symmetry
aligned along the 3-axis and with angular momentum $K$ about this
axis.  Let $\mathcal{O}_K$ be the coadjoint orbit containing
$\rho_0^{(K)}$.  When $K\neq 0$, the density $\rho_0^{(K)}$ is no
longer invariant under rotations through $\pi$ about an axis
perpendicular to the symmetry axis and $\mathcal{O}_K$ becomes
symplectomorphic to $T^*(SO(2) \backslash SO(3))$; as a manifold,
$\mathcal{O}_K$ remains four-dimensional.

\subsubsection{The classical dynamics of a symmetric top}

The classical dynamics of a symmetric top illustrate the advantages of
working algebraically with observables rather than coordinates and of
considering the component of angular momentum $K$ about the symmetry
axis as a gauge degree of freedom.

Suppose the classical Hamiltonian for a symmetric top is given by the
standard function
\begin{equation}
  \label{eq:RotorHamiltonianFunction}
  \mathcal{H} = \frac{1}{2} \sum_{mn} \mathcal{L}_m \Im^{-1}_{mn}
  \mathcal{L}_n \,. 
\end{equation} 
Because $\mathcal{H}$ is rotationally invariant, the square of the
angular momentum $\mathcal{L}^2$ is a constant of the motion.  And,
for a symmetric top, the component $K$ of the angular momentum along
the symmetry axis is also a constant of the motion.  Thus, in the
principal axes frame of the rotor, the Hamiltonian becomes
\begin{equation}
  \mathcal{H} = \displaystyle \frac{1}{2{\Im}_1}
  (\mathcal{L}_1^2
  +\mathcal{L}_2^2) + \frac{1}{2{\Im}_3} K^2 = 
   \frac{1}{2{\Im}_1}
   \mathcal{L}^2 + {\rm constant} \,,
\end{equation}
where
\begin{equation}
  \mathcal{L}^2 =  \mathcal{L}_1^2+ \mathcal{L}_2^2+ \mathcal{L}_3^2 \,.
\end{equation}
Although derived in the principal axes frame, these expressions of
$\mathcal{H}$ and $\mathcal{L}^2$ are valid in any reference frame,
albeit with $\Im_1$ and $K$ regarded as numerical constants.

Now, because the phase space of a symmetric top is of dimension four,
the motion of the rotor is characterized by the time evolution of any
four linearly-independent observables, e.g., the components
$\{\mathcal{L}_1,\mathcal{L}_2,\Im_{13},\Im_{23} \}$ of $\mathcal{L}$
and $\Im$ relative to the space-fixed axes.  The time evolution of
these observables is then given by solution of the equations of motion
\begin{equation}
    \dot{\Im}_{i3} = \frac{1}{2{\Im}_1} 
     \{ \Im_{i3}, \mathcal{L}^2 \}\,, \qquad 
     \dot{\mathcal{L}}_i = \frac{1}{2{\Im}_1}\{
    \mathcal{L}_i, \mathcal{L}^2  \} = 0 \,, \quad i=1,2.
\end{equation}
As expected, these equations confirm that each component of the
angular momentum is conserved.

Suppose that the angular momentum has magnitude $L$ and is aligned
along the space-fixed 3-axis.  Then the time evolution of the top is
given by
\begin{equation}
  \dot{\Im}_{13} = - \frac{L}{\Im_1} \Im_{23} \, ,\qquad
  \dot{\Im}_{23} = \frac{L}{\Im_1} \Im_{13} \, .
\end{equation}
These are the equations of a simple two-dimensional harmonic
oscillator of frequency $L/\Im_1$.  Thus, the top precesses about the
3-axis with this angular frequency.  Note, however, that if the
angular momentum lies along the symmetry axes of the symmetric top,
then the symmetry axis coincides with the space-fixed 3-axis.  And,
since the symmetry axis is a principal axis of the inertia tensor, it
then follows that $\Im_{12}= \Im_{13}=0$ and the top simply spins in
the expected way, without precession, with angular momentum $K=L$
about its symmetry axis.

\subsubsection{Semi-classical representations of the symmetric top}

The intrinsic degrees of freedom of a symmetric top are quantized in a
semi-classical representation by replacing the classical phase space
$[\mathbb{R}^4]D_\infty \backslash [\mathbb{R}^6]SO(3)$ $\simeq T^*
(D_{\infty}\backslash SO(3))$ by a fibre bundle associated to the
principal $ [\mathbb{R}^6]SO(3) \to [\mathbb{R}^4]D_\infty \backslash
[\mathbb{R}^6]SO(3)$ bundle by a unitary irrep $M$ of the isotropy
subgroup $[\mathbb{R}^4]D_\infty$.  Such semi-classical
representations can be derived from an abstract unitary representation
$T$ of the RMG on a Hilbert space $\mathbb{H}$ as follows.

Let $|\xi_K\rangle = |K\rangle \in\mathbb{H}$ be a normalized state
that satisfies equation~(\ref{eq:RMAModifiedDensityOrigin}) with
$\Im_1=\Im_2\not=\Im_3$, and let $|\xi_{\bar{K}}\rangle$ be defined by
\begin{equation}
  \label{eq:RelationKandKbar}
  |\xi_{\bar K}\rangle = |\bar K\rangle = T(\rme^{\frac{\rmi}{\hbar}\pi
  L_2})|K\rangle \, .  
\end{equation}
Let
\begin{equation}
   E = |K\rangle\langle \xi_K| + |\bar K\rangle\langle \xi_{\bar K}|
\end{equation}
be the natural embedding of the subspace $U$, spanned by the states
$\{ |\xi_K\rangle, |\xi_{\bar K}\rangle\}$, in $\mathbb{H}$, and let
\begin{equation}
   \Pi = |\xi_K\rangle\langle K| + |\xi_{\bar K}\rangle\langle {\bar K}|
\end{equation}
be the corresponding $\mathbb{H}\to U$ projection operator.  Together,
$E$ and $\Pi$ define a semi-classical density $\hat{\rho}^{(K)}(A) =
\Pi T(A) E$ that (for $K\not= 1/2$) satisfies
\begin{equation}
  \label{eq:RMASemiClassicalDensity}
  \eqalign{
  \hat\rho^{(K)}(L_i) &= 0 \, , \quad
  i=1,2\, , \\ 
  \hat\rho^{(K)}(L_3) &= K\Bigl( |\xi_K\rangle \langle \xi_K| -      
  |\xi_{\bar K}\rangle \langle\xi_{\bar K}| \Bigr) 
  \equiv \hat{\mathcal{S}} \, , \\
  \hat\rho^{(K)}(I_{ij}) &= \delta_{ij}{\Im}_i
  \Bigl( |\xi_K\rangle \langle \xi_K| +      
  |\xi_{\bar K}\rangle \langle\xi_{\bar K}| \Bigr) =
  \delta_{ij}{\Im}_i \hat{\mathcal{I}} \, ,
  }
\end{equation}
with $\hat{\mathcal{I}}$ the identity operator on $U$.

The subalgebra $\mathfrak{h}$ of the RMA, $\mathfrak{g}
= [\mathbb{R}^6]so(3)$, defined by
\begin{equation} 
  \label{eq:7.}
  \mathfrak{h} = \{ A\in \mathfrak{g} |\, \hat\rho ([A,B]) = 
  [\hat\rho(A),\hat\rho(B)] \,,\;\forall\, B\in \mathfrak{g}\} \,,
\end{equation}
is the Lie algebra $[\mathbb{R}^4]so(2)$.  Moreover, the restriction
of $\hat\rho$ to $\mathfrak{h}\subset \mathfrak{g}$ is a reducible
representation $M$ for which
\begin{equation}
  \label{eq:EigenstatesOfR3}
  \eqalign{
  M(L_3)|\xi_K\rangle &=K |\xi_K\rangle \, , \qquad
  M(L_3)|\xi_{\bar K}\rangle = -K |\xi_{\bar K}\rangle\, , \\
  M(I_{ii})|\xi_K\rangle &= {\Im}_i |\xi_K\rangle \, , \qquad
  M(I_{ii})|\xi_{\bar K}\rangle = {\Im}_i |\xi_{\bar K}\rangle
  \, , \quad i=1,2,3,\\
  M(I_{12})|\xi_K\rangle &= M(I_{12})|\xi_{\bar K}\rangle =0 \,.
  }
\end{equation}
Note that, unless $U\subset \mathbb{H}$ happens to be an
$\mathfrak{h}$-invariant subspace, this representation of
$\mathfrak{h}$ is not a subrepresentation of the restriction of $T$ to
$\mathfrak{h}\subset \mathfrak{g}$.  It is an example of an embedded
representation, as discussed in section~\ref{sec:Partial}.  Nevertheless,
it integrates to a reducible (and generally projective) unitary irrep
$M$ of $[\mathbb{R}^4]SO(2)$ which extends to an irreducible unitary
irrep of $[\mathbb{R}^4]D_\infty$ with, for example,
\begin{equation}
  \eqalign{
  M(\rme^{-\frac{\rmi}{\hbar} \theta L_3}) |\xi_K\rangle &=
  \rme^{-\rmi K\theta} |\xi_K\rangle \,,\\
  M(\rme^{-\frac{\rmi}{\hbar} \theta L_3}) |\xi_{\bar K}\rangle &=
  \rme^{\rmi K\theta} |\xi_{\bar K}\rangle\,,\\
  M(\rme^{-\frac{\rmi}{\hbar} \pi L_2}) |\xi_K\rangle &= (-1)^{2K} |\xi_{\bar
  K}\rangle \,,\\
  M(\rme^{-\frac{\rmi}{\hbar} \pi L_2}) |\xi_{\bar K}\rangle &=  |\xi_K\rangle
  \,.
  }
\end{equation} 

The operator $\hat\rho$ also defines a semi-classical representation
of any element $A$ in the RMA by an operator valued function
$\hat{\mathcal{A}}$ over the RMG with values
\begin{equation}
  \hat{\mathcal{A}}(g) = \hat\rho (A(g)) \,, \quad g\in
  [\mathbb{R}^6]SO(3) \,,
\end{equation}
where $A(g) = {\rm Ad}_g(A)$.  These functions satisfy the
$[\mathbb{R}^4]D_\infty$-equivariance condition
\begin{equation}
  \hat\rho_{hg}  =M(h) \hat\rho_gM(h^{-1})
  \,, \quad \forall\ h\in [\mathbb{R}^4]D_\infty\,,
\end{equation}
and have Poisson brackets
\begin{equation}
  \{\hat{\mathcal{A}},\hat{\mathcal{B}}\}(g) = - \frac{\rmi}{\hbar}
  \hat\rho ([A(g),B(g)]) \, , \quad \forall\ g\in [\mathbb{R}^6]SO(3) \,.
\end{equation}

\subsubsection{Quantization of a symmetric top}

There is a natural polarization for any cotangent bundle and, as a
result, the full quantization of a rotor is simpler than
prequantization.  We therefore bypass prequantization and proceed
directly to quantization by constructing an appropriate unitary irrep
of the RMG.  The natural polarization for the symmetric top is defined
by starting with a representation $M$ of the isotropy subgroup
$[\mathbb{R}^4]D_\infty$ for the phase space of a symmetric top and
extending it to a representation $\tilde M$ of
$[\mathbb{R}^6]D_\infty$. Such a representation is defined as
\begin{equation}
  \label{eq:RepOfPolarization}
  \tilde M (Q,\omega) =
  \rme^{-\frac{\rmi}{\hbar}Q\cdot \overline{\Im}} M(\omega) \, ,
  \quad  Q\in \mathbb{R}^6\,, \; \omega\in D_\infty \,,
\end{equation}
where $\overline{\Im}$ is the diagonal matrix whose entries are the
principal moments of inertia $(\Im_1,\Im_2,\Im_3)$ of the rotor, and
$Q\cdot \overline{\Im} = \sum_{ij} Q_{ij} \overline{\Im}_{ij}
=\sum_{i} Q_{ii}{\Im}_i $.

Note, however, that for the semi-classical representation defined by
$M$ to be quantizable, it is necessary that $2K$ should be an integer.
Otherwise the representation of $SO(2)$ labelled by $K$ will not be
contained in any representation of $SO(3)$.  If $2K$ is odd, then $M$
is contained in a projective (spinor) representation of $SO(3)$, i.e.,
a true representation of $SU(2)$, the double cover of $SO(3)$. Thus, to
avoid the subtleties associated with projective representations, it
will be convenient in the following to regard $\tilde M$ as a true
irrep of $[\mathbb{R}^4]D^d_\infty$, the double cover of
$[\mathbb{R}^4]D_\infty$, and require that it be contained in some
unitary representation of $[\mathbb{R}^4]SU(2)$.

Let $U$ be the carrier space for the irrep $\tilde M$ of
$[\mathbb{R}^6]D^d_\infty$.  Now, we no longer require $U$ to be a
subspace of the Hilbert space $\mathbb{H}$ for the abstract
representation $T$ of the RMG.  Instead, an irrep of the RMG is
induced in VCS theory by defining an
$[\mathbb{R}^6]D^d_\infty$-intertwining operator $\Pi :\mathbb{H}_D
\to U$ from a suitably defined dense subspace
$\mathbb{H}_D\subset\mathbb{H}$ to $U$, such that
\begin{equation}
  \label{eq:RotorIntOp}
  \Pi T(Q,\omega) = \rme^{-\frac{\rmi}{\hbar}Q\cdot \overline{\Im}}M(\omega)
  \Pi \,, 
  \quad Q\in \mathbb{R}^6\,, \; \omega\in D^d_\infty \,. 
\end{equation}
VCS wave functions are then defined initially as vector-valued
functions over $[\mathbb{R}^6] SU(2)$ with values in $U$ given by
\begin{equation}
  \Psi(Q,\Omega) = \Pi T(Q,\Omega) |\Psi\rangle \,,\quad
  |\Psi\rangle \in \mathbb{H}_D \,.
\end{equation}
Because of the constraint condition (\ref{eq:RotorIntOp}), these
functions satisfy
\begin{equation}
  \Psi(Q,\Omega) =  \rme^{-\frac{\rmi}{\hbar}Q\cdot \overline{\Im}}
  \Pi R(\Omega) |\Psi\rangle \,,
  \quad Q\in \mathbb{R}^6\,, \; \Omega\in SU(2)\,,
\end{equation}
where $R(\Omega) = T(0,\Omega)$ is the restriction of the
representation $T$ to $SU(2)$.  Thus, it is sufficient to define VCS
wave functions as the vector-valued functions over $SU(2)$
\begin{equation}
  \psi(\Omega) = \Pi R(\Omega) |\Psi\rangle\,,\quad
  |\Psi\rangle \in \mathbb{H}_D \,, \, \Omega \in SU(2) \, ,
\end{equation}
which satisfy the condition
\begin{equation}
  \psi(\omega\Omega) = M(\omega) \psi(\Omega) \,,\quad \forall\ \omega\in
  D^d_\infty\,.
\end{equation}
The VCS representation of the RMG is now defined on these wave
functions by
\begin{equation}
  \eqalign{
  [\Gamma(Q,\Omega)\psi](\Omega') &= \Pi R(\Omega') T(Q,\Omega)
  |\Psi\rangle \\
  &= \Pi T(\Omega' Q \tilde \Omega',\Omega'\Omega)
  |\Psi\rangle }  
\end{equation}
which gives
\begin{equation}
  \label{eq:7.VCSrotorrep}
  [\Gamma(Q,\Omega)\psi](\Omega') =
  \rme^{-\frac{\rmi}{\hbar}(\Omega'Q\tilde\Omega')\cdot
  \overline{\Im}} \Psi(\Omega'\Omega) \,.
\end{equation} 

An explicit construction of the Hilbert space for this VCS
representation is presented as follows.  First observe from
equation~(\ref{eq:7.VCSrotorrep}) that a reducible representation $T$
of the RMG is defined on the Hilbert space $\mathbb{H} =
\mathcal{L}^2(SU(2))$ by
\begin{equation}
  [T(Q,\Omega)\psi](\Omega') =
  \rme^{-\frac{\rmi}{\hbar}(\Omega'Q\tilde\Omega')\cdot
  \overline{\Im}} \Psi(\Omega'\Omega) \,.
\end{equation}
Now, by the Peter-Weyl theorem, an orthonormal basis for
$\mathcal{L}^2(SU(2))$ is given by the $SU(2)$ Wigner functions
\begin{equation}
  \Phi_{NJM} = \sqrt{\frac{2J+1}{8\pi^2}}\, \mathcal{D}^J_{NM} \,,
\end{equation}
where $2J$ is a positive or zero integer and $M$ and $N$ run from $-J$
to $+J$ in integer steps.  Let $\{ |NJM\rangle \}$ denote the vector
in $\mathbb{H}$ with wave function $\Phi_{NJM}$ and let $\mathbb{H}_D$
denote the dense subspace of finite linear combinations of these basis
vectors.  Now let $\langle K|$ and $\langle \bar K|$ denote the
functionals on $\mathbb{H}_D$ for which
\begin{equation}
  \eqalign{
  \langle K| NJM\rangle =
  \sqrt{\frac{2J+1}{8\pi^2}}\,\delta_{NK}\delta_{MK} \,,\\
  \langle \bar K| NJM\rangle = (-1)^{J+K}
  \sqrt{\frac{2J+1}{8\pi^2}}\,\delta_{NK}\delta_{M,-K} \,.
  }
\end{equation}
Let $\Pi$ denote the operator
\begin{equation}
  \Pi = \frac{1}{\sqrt{2}} \Bigl( \xi_K\langle K| + \xi_{\bar K} \langle
  \bar K| \Bigr)
\end{equation}
that maps $\mathbb{H}_D\to U$, where $\{\xi_K,\xi_{\bar K}\}$ is the
basis for $U$ as defined above with $2K$ a fixed positive integer.
This operator satisfies the intertwining condition
\begin{equation}
  \Pi R(\omega) = M(\omega) \Pi \,, \quad \forall\ \omega \in
  D^d_\infty \,,
\end{equation}
and defines a basis $\{\psi_{KJM}\}$ for a Hilbert space
$\mathcal{H}^K$ of coherent state wave functions, having values
\begin{equation}
  \eqalign{
  \psi_{KJM}(\Omega) &= \Pi R(\Omega)|KJM\rangle \\ 
  &= \sqrt{\frac{2J+1}{16\pi^2}} \left[ \xi_K\mathcal{D}^J_{KM}(\Omega)
  +(-1)^{J+K} \xi_{\bar K} \mathcal{D}^J_{-K,M}(\Omega)  \right]\,.
  }
  \label{eq:7.168}
\end{equation}
This basis is seen to be orthornormal relative to the natural
$U\otimes \mathcal{L}^2(SU(2))$ inner product.  It is the standard
basis of rotor model wave functions used in nuclear
physics~\cite{bohr,rowebook,bohrmottelson}.

The map $\mathbb{H}\to \mathcal{H}^K$, defined by
equation~(\ref{eq:7.168}), shows that $\mathcal{H}^K$ is isomorphic to
a subspace of $\mathbb{H}$.  From the theory of induced
representations, it is known that this subspace is irreducible.  Thus,
the irrep $\tilde M$ of the subgroup $[\mathbb{R}^6]D^6_\infty$
uniquely defines an irreducible representation of the RMG and its Lie
algebra RMA and, hence, a quantization of the symmetric top model.

\section{Concluding remarks}

Coherent state representation theory has its most general expression
in vector coherent state (VCS) theory.  This theory is now highly
developed as a practical theory of induced representations.  It
encompasses virtually all the standard inducing constructions.  In
addition, it facilitates the construction of orthonormal bases for
irreducible representations and provides practical algorithms for the
computation of the matrix coefficients for the irreps of model
spectrum generating algebras.  By having the flexibility to induce
irreps of a group $G$ from a multidimensional irrep of a subgroup
$H\subset G$, VCS theory has a huge practical advantage over its
scalar counterpart.  It has been used to construct irreps of
representative examples of all the classical Lie algebras and has been
applied widely to models in nuclear physics (cf.\ references cited in
\cite{rowe96} and \cite{rowe91}).

The relationship of geometric quantization to the theory of induced
representations is surely well understood by experts in the two
fields.  However, the new insights and simplifications that can be
brought to the practical application of both theories by VCS theory is
not known.  We hope to have shown in this paper that, by understanding
the relationships between the three theories when they are expressed
in a common language, it becomes possible to exploit their
complementary features to greatest advantage.

Already some new perspectives and new approaches to old physical
problems are suggested by the unified approach to quantization
presented here.  An important advance in modern physics has been the
development of abelian and non-abelian gauge theory.  It has long
been known that (often hidden) intrinsic motions can have a profound
effect on the dynamics of a system.  A well-known example of this is
the precessional motion of a symmetric top that is spinning in a way
that may not be directly observable about its symmetry axis.  

The VCS methods outlined in this paper suggest ways to select
physically and mathematically relevant intrinsic degrees of freedom
and express their influence on the complementary extrinsic dynamics in
terms of gauge potentials.  For example, it might be appropriate to
regard the fast and slow degrees of freedom of a many-body system,
for which the Born-Oppenheimer approximation~\cite{BO} applies, as
being intrinsic and collective, respectively.  A model description of
the scattering of two nuclei for which the motions of their centres of
mass are {\em adiabatic} in comparison to their intrinsic degrees of
freedom would be a candidate for such a separation.  Quantum optics
also provides examples such as parametric down conversion (PDC) that
possess useful algebraic models~\cite{PDC}; in the standard
description of PDC, strong coherent beams are treated classically
while down-converted photon pairs must be described quantally.  In
physical systems where some degrees of freedom tend to behave in a
manifestly quantal way while others are essentially classical, the
proposals given in Sec.~\ref{sec:Partial} for deriving partial
quantizations have the potential for providing systematic ways of
modelling such systems.  In particular, methods are given for
constructing semi-classical models in which the intrinsic (gauge)
degrees of freedom are quantized but the extrinsic dynamics are
treated classically.

A problem of considerable interest is the description of vortices in
quantum fluids.  This problem has been related to the notoriously
difficult task of constructing the unitary irreps of
infinite-dimensional groups of diffeomorphisms \cite{diffeo}.  The
potential relevance of the methods proposed in this paper to this
problem are indicated by the following observations.  A model of
linear hydrodynamic flows in nuclei (the so-called CM(3) model) has
been quantized both by induced representation methods \cite{RR76} and
geometric quantization \cite{RI79}.  These quantizations are
characterized by quantized vortex spins which are naturally regarded
as intrinsic $SU(2)$ degrees of freedom.  The second observation is
that VCS methods have been successfully applied to the
infinite-dimensional affine Lie algebra $\widehat{sl}(2)$
\cite{Zhang}.

\ack

We thank C Bahri for useful discussions.  SDB acknowledges the support
of a Macquarie University Research Fellowship.  This paper was
supported in part by NSERC of Canada.
 
\appendix 

\section{The covariant derivative and curvature tensor}
\label{sec:AppendixA}

\emph{Claim:} Let $B$ be a vector bundle with typical fibre $U$
associated to a principal $G\to H\backslash G$ bundle by a unitary
representation $M$ of $H \subset G$. Let
$\hat \rho$ be an $H$-equivariant $\mathfrak{g}\to GL(U)$ map having
the property that it maps the subalgebra $\mathfrak{h}\subset
\mathfrak{g}$ to the representation $M$, i.e., $\hat \rho (A) = M(A)$
for $A\in \mathfrak{h}$ (cf., text for details).   Define
\begin{equation} 
  \label{eq:CovDeriv}
  \rmi\hbar [\nabla_{A} \Psi](g)  = \Psi(A(g)g) - \hat\rho(A(g))\Psi(g)\,,
\end{equation}
where $A(g) = {\rm Ad}_g(A)$, $\Psi(Ag)$ is defined for any $A\in
\mathfrak{g}$ by equation~(\ref{eq:4.psiAg}), and $\Psi$ is any
section of $B$, i.e., it satisfies the identity
\begin{equation}
  \Psi(hg) = M(h) \Psi(g) \,,\quad \forall\ h\in H\,.
\end{equation}
Then $\nabla_A$ is identical to
\begin{equation}
  \nabla_{X_{\hat{\mathcal{A}}}} =X_{\hat{\mathcal{A}}} +
  \frac{\rmi}{\hbar} \hat\theta(X_{\hat{\mathcal{A}}}) \,,
\end{equation}
the covariant derivative in the direction of the Hamiltonian vector
field $X_{\hat{\mathcal{A}}}$ over $H\backslash G$ generated by the
vector-valued function $\hat{\mathcal{A}}(g) = \hat\rho(A(g))$, where
$\hat\theta$ is a symplectic connection (one-form) for $B$.  \medskip

\noindent\emph{Proof:} A choice of gauge is defined by the expansion
\begin{equation}
  A(g)= \sum_i A^i(g) A_i + \sum_\nu A^\nu(g) A_\nu \, ,
\end{equation}
where $\{ A_i\}$ is a basis for $\mathfrak{h}$ and $\{ A_\nu\}$
completes a basis for $\mathfrak{g}$.  Using the identity
$\Psi(A_i g) = \hat\rho(A_i)\Psi(g)$, the definition
(\ref{eq:CovDeriv}) gives 
\begin{equation} 
  \label{eq:A.nablaA} 
  \rmi\hbar [\nabla_{A} \Psi](g)  =
  \sum_\nu A^\nu (g) \bigl( \Psi(A_\nu g) - \hat\rho (A_\nu) \Psi(g)
  \bigr) \,.
\end{equation}
Now, if $g(x)= \rme^{X(x)}g$, with  $X(x) = -\frac{\rmi}{\hbar} \sum_\mu
x^\mu A_\mu$, then as shown in the appendix to~\cite{scalar}, 
\begin{equation}
  \rmi\hbar\, \displaystyle\frac{\partial }{\partial x^\nu} \Psi(g(x)) =
  \Psi(A_\nu(x) g(x)) \,,
\end{equation}
where
\begin{equation} 
  \label{eq:A.Anu}
  \eqalign{
  A_\nu(x) &= -\rmi\hbar \, \rme^{X(x)} \frac{\partial}{\partial x^\nu}
  \rme^{-X(x)} \\
  &=A_\nu + \frac{1}{2!} [X(x),A_\nu] + 
  \frac{1}{3!} [X(x),[X(x),A_\nu]]+ \cdots \,.
  }
\end{equation}
Therefore, if $A_\nu(x)$ is expanded
\begin{equation} 
  \label{eq:AnuExpansion}
  A_\nu(x) = \sum_\mu \Lambda_\nu^\mu(x) A_\mu + \sum_i
  \lambda_\nu^i(x) A_i \,, 
\end{equation}
then
\begin{equation} 
  \Psi(A_\nu g(x)) = \rmi\hbar\, \sum_\mu \bar\Lambda^\mu_\nu(x)
  \Bigl(\frac{\partial }{\partial x^\mu} 
  + \frac{\rmi}{\hbar}\lambda^i_\mu(x)
  \hat \rho(A_i)\Bigr)\Psi(g(x)) \,,
\end{equation}
where $\bar\Lambda$ is the inverse of the matrix $\Lambda$.  It
follows from equation~(\ref{eq:A.nablaA}) that
\begin{equation}  
  [\nabla_{A} \Psi](g(x))  =
  \sum_\nu A^\nu (g(x)) \bar\Lambda^\mu_\nu(x) \Bigl( 
  \frac{\partial }{\partial x^\mu} + \frac{\rmi}{\hbar}\hat\theta_\mu(x)
  \Bigr)\Psi(g(x)) \,,
\end{equation}
where
\begin{equation} 
  \label{eq:A.theta}
  \hat\theta_\mu(x) = \sum_\nu \Lambda^\nu_\mu(x)
  \hat \rho(A_\nu) + \sum_i\lambda^i_\mu(x)
  \hat \rho(A_i)  = \hat\rho(A_\mu(x)) \,.
\end{equation}
Thus, if we regard $\hat\theta_\mu(x)$ as the component
$\hat\theta_{g(x)}\big(\partial /\partial x^\mu \big)$ of a one-form
$\hat\theta$, defined at $g(x)$ by 
\begin{equation}
  \hat\theta_{g(x)} = \sum_\mu \hat\theta_\mu(x) \rmd x^\mu \,,
\end{equation}
and define the Hamiltonian vector field $X_{\hat{\mathcal{A}}}$ by
\begin{equation} 
  \label{eq:A.vecfield}
  [X_{\hat{\mathcal{A}}}\Psi](g(x)) =  \sum_\nu A^\nu (g)
  \bar\Lambda^\mu_\nu(x)  \frac{\partial }{\partial x^\mu} \Psi(g(x))\,,
\end{equation}
then
\begin{equation}  
  \nabla_{A} = X_{\hat{\mathcal{A}}}
  +\frac{\rmi}{\hbar}\hat\theta(X_{\hat{\mathcal{A}}}) \, ,
\end{equation}
as claimed.  

To check that $\hat \theta$ is a symplectic connection, we now derive
the curvature of the connection one-form $\hat \theta$.  Consider
first the standard exterior derivative of $\hat\theta$ given by
\begin{equation}
  \rmd\hat\theta_{g(x)} = \sum _{\mu\nu}
  \frac{\partial\hat\theta_\nu (x)}{\partial x^\mu}\, 
  \rmd x^\mu \wedge \rmd x^\nu \,.
\end{equation}
From the definition (\ref{eq:A.theta}) of $\hat\theta_\nu (x)$, and
with $A_\nu(x)$ expressed by equation~(\ref{eq:A.Anu}),
\begin{equation}
  \frac{\partial\hat\theta_\nu (x)}{\partial x^\mu}
  =-\rmi\hbar \,\frac{\partial}{\partial x^\mu}  \hat\rho
  \Bigl( \rme^{X(x)} \frac{\partial}{\partial x^\nu} \rme^{-X(x)}\Bigr) .
\end{equation}
Then, with the help of the identities (see~\cite{scalar})
\begin{eqnarray}
  \rmi\hbar\, \displaystyle\frac{\partial \rme^{X(x)}}{\partial x^\nu} =
  A_\nu(x) \rme^{X(x)} = \rme^{X(x)}A_\nu(-x)\,, \\
  \rmi\hbar\, \displaystyle\frac{\partial \rme^{-X(x)}}{\partial x^\nu} =-
  A_\nu(-x) \rme^{-X(x)} = -\rme^{-X(x)}A_\nu(x)\,, 
\end{eqnarray}
we obtain
\begin{equation}
  \rmd\hat\theta_{g(x)}\left( \frac{\partial}{\partial x^\mu},
  \frac{\partial}{\partial x^\mu} \right)
  = \frac{\partial\hat\theta_\nu (x)}{\partial x^\mu}
  =-\frac{\rmi}{\hbar} \, \hat\rho \left([ A_\mu(x),A_\nu(x)]\right) . 
\end{equation}
Thus, from the expansion of $A_\nu(x)$ given by
equation~(\ref{eq:AnuExpansion}), and recalling that
\begin{equation}
  \hat\rho ([A,B]) = [\hat\rho(A), \hat\rho(B)] \,,\quad A\in
  \mathfrak{h}\,,\ B\in \mathfrak{g} \,,
\end{equation}
we derive
\begin{equation}
  \rmd\hat\theta_{g(x)}\left( \frac{\partial}{\partial x^\mu},
  \frac{\partial}{\partial x^\mu} \right)
  = \sum_{\mu'\nu'} \Lambda^{\mu'}_\mu (x) \hat\Omega_{\mu'\nu'}
  \Lambda^{\nu'}_\nu (x) + [\hat\theta_\mu(x),\hat\theta_\nu(x)] .
\end{equation}
with
\begin{equation}
  \hat\Omega_{\mu\nu} = - \frac{\rmi}{\hbar}\Bigl( \hat\rho([A_\mu,
  A_\nu]) - [\hat\rho(A_\mu),\hat\rho(A_\nu)]\Bigr) \, .
\end{equation}
It follows that, for the vector fields defined by
equation~(\ref{eq:A.vecfield}),
\begin{equation}
  \rmd\hat\theta_{g(x)}\left( X_{\hat{\mathcal{A}}},
  X_{\hat{\mathcal{B}}}\right)
  = \sum_{\mu\nu} A^\mu(g(x)) \hat\Omega_{\mu\nu} B^\nu(g(x)) + 
  \big[\hat\theta_{g(x)} \big( X_{\hat{\mathcal{A}}}\big),\hat\theta_{g(x)}
  \big( X_{\hat{\mathcal{B}}}\big)\big] \,.
\end{equation}
Hence we derive the general expression for the curvature tensor
\begin{equation}
  \hat\Omega\left( X_{\hat{\mathcal{A}}},  X_{\hat{\mathcal{B}}}\right) =
  \rmd\hat\theta\left( X_{\hat{\mathcal{A}}}, X_{\hat{\mathcal{B}}}\right) -
  \big[\hat\theta \big( X_{\hat{\mathcal{A}}}\big),\hat\theta
  \big( X_{\hat{\mathcal{B}}}\big)\big] \,.
\end{equation}
\hfill \emph{QED}

\end{document}